\documentclass[a4paper,11pt]{article}
\pdfoutput=1 

\usepackage{jheppub} 

\usepackage[T1]{fontenc} 
\usepackage{dsfont}

\usepackage{Settings}
\usepackage{blkarray}
\usepackage{todonotes}
\usepackage{subcaption}
\usepackage{tabularx}
\usepackage{booktabs}
\usepackage{mathrsfs}

\usepackage{accents}

\newcommand{\badat}{\begin{alignedat}}
 \newcommand{\eadat}{\end{alignedat}}

\usepackage[colorlinks = true,
            linkcolor = blue,
            urlcolor  = blue,
            citecolor = blue,
            anchorcolor = blue]{hyperref}

\usepackage{comment}

\preprint{
\vspace{-24pt}
\begin{flushright}
CPHT-RR046.062022\\
HU-EP-22/26\\
SAGEX-22-24-E\\
TCD 22-04
\end{flushright}
}

\DeclareMathOperator\arctanh{arctanh}
\newcommand*\DAlambert{\mathop{}\!\mathbin\Box}
\usepackage{blindtext}%
\usepackage{etoolbox}%
\let\oldbibliography\bibliography
\renewcommand{\bibliography}[1]{{%
  \let\chapter\section
  \oldbibliography{#1}}}

\title{\boldmath Celestial holography on Kerr-Schild backgrounds}

\author[1,2]{Riccardo Gonzo,}
\author[1,3]{Tristan McLoughlin,}
\author[4]{Andrea Puhm}

\affiliation[1]{School of Mathematics \& Hamilton Mathematics Institute, Trinity College Dublin, College Green, Dublin 2, Ireland}
\affiliation[2]{Higgs Centre for Theoretical Physics, School of Physics and Astronomy, The University of Edinburgh, EH9 3FD, Scotland}
\affiliation[3]{Institut f\"ur Physik und IRIS Adlershof, Humboldt-Universit\"at zu Berlin, \\
  Zum Gro{\ss}en Windkanal 2, D-12489 Berlin, Germany}
\affiliation[4]{CPHT, CNRS, Ecole Polytechnique, IP Paris, F-91128 Palaiseau, France}

\emailAdd{gonzo@maths.tcd.ie,tristan@maths.tcd.ie,andrea.puhm@polytechnique.edu}

\abstract{
We explore the celestial holography proposal for non-trivial asymptotically flat backgrounds including the Coulomb field of a static and spinning point charge, their gravitational counterparts described by the Schwarzschild and Kerr metrics, as well as the Aichelburg-Sexl shockwave and spinning shockwave geometries and their electromagnetic cousins. We compute celestial two-point amplitudes on these Kerr-Schild type backgrounds which have the desirable feature, due to the presence of an external source, that they are non-vanishing for general operator positions and are not constrained by the kinematic delta
functions of flat space celestial CFT correlators. Of particular interest is the case of shockwave backgrounds where the two-point scattering amplitude of massless scalars can be interpreted as a standard CFT three-point correlator between two massless asymptotic states and a conformal primary shockwave operator. We furthermore show that the boundary on-shell action for general backgrounds becomes the generating functional for tree-level correlation functions in celestial CFT. Finally, we derive (conformal) Faddeev-Kulish dressings for particle-like backgrounds which remove all infrared divergent terms in the two-point functions to all orders in perturbation theory.

}

\begin{document} 
 
\maketitle
\flushbottom

\section{Introduction}

In the past few decades ever more sophisticated methods have been developed to probe properties of quantum and gravitational systems. 
Key in this endeavor are scattering amplitudes whose remarkable simplicity and hidden structure, emerging after extensive calculations, fuel the search for new conceptual formulations.
The prospect that physical theories can be strongly constrained using general principles alone, and sometimes even solved or bootstrapped, raises the hope for a non-perturbative understanding of the S-matrix.
The other major tool that has emerged is holography which offers a non-perturbative definition of bulk quantum-gravitational physics through a conjectured duality with a quantum field theory at the boundary of the spacetime. 

A concrete realization of the holographic principle is achieved in asymptotically Anti-de Sitter (AdS) spacetimes whose dual conformal field theory (CFT) lives on the timelike co-dimension one boundary, but it is far from obvious if it applies to more general spacetimes. Asymptotically flat (AF) four-dimensional spacetimes are a good approximation to many physical processes in the universe. The S-matrix, which is the basic observable in  such spacetimes, has an obvious  holographic flavor in that it is defined by a set of on-shell data at the asymptotic boundary of the spacetime. 
Upon a change of basis from asymptotic momentum to boost eigenstates, scattering amplitudes in four spacetime dimensions transform as conformal correlators~\cite{deBoer:2003vf,He:2015zea,Cheung:2016iub,Pasterski:2016qvg,Cardona:2017keg} on the asymptotic two-sphere at the conformal boundary of the AF spacetime where the Lorentz group acts as the Euclidean global conformal group. In a quantum theory of gravity the latter gets enhanced to the full local conformal group~\cite{Cachazo:2014fwa,Kapec:2014opa,Kapec:2016jld}.

These insights have led to the celestial holography proposal which pursues a potential duality between quantum gravity in AF spacetimes and a celestial conformal field theory (CCFT) on the co-dimension two celestial sphere.
A number of entries in the celestial holography dictionary have by now been established, though primarily for flat space scattering amplitudes\footnote{Some exceptions are discussed in the recent works~\cite{Costello:2022wso,Costello:2022upu,Fan:2022vbz,Casali:2022fro}.} and with a focus on the CCFT properties of scattering processes involving massless particles such as identifying the celestial avatars of asymptotic symmetries, soft theorems and other universal aspects of the S-matrix\footnote{See the recent reviews and lecture notes~\cite{Strominger:2017zoo,Raclariu:2021zjz,Pasterski:2021rjz,Pasterski:2021raf,McLoughlin:2022ljp} and the references therein.}.
A litmus test of celestial holography is whether it can account for non-perturbative physics such as the formation and evaporation of black holes from the perspective of the boundary theory. This is a highly challenging open question. A much simpler problem, yet a crucial stepping stone, is to describe non-trivial AF backgrounds in CCFT and their effects on celestial scattering amplitudes. In this work we take some intitial steps in this direction. Along the way, we will draw attention to expectations from AdS/CFT and differences that arise in AF spacetimes.

Focusing on the scattering of massless probe particles, a Mellin transform in the external particle energies takes momentum space amplitudes prepared with plane waves to celestial amplitudes which are prepared with conformal primary wavefunctions~\cite{Pasterski:2017kqt}. 
The primary operators appearing in these CCFT correlators~\cite{Pasterski:2016qvg} can be constructed from an ``extrapolate'' dictionary that maps bulk fields to boundary operators~\cite{Donnay:2020guq,Pasterski:2021dqe}. This resonates with the way boundary correlators in AdS/CFT are extracted from bulk correlators that are pushed to the timelike AdS boundary~\cite{Susskind:1998dq, Banks:1998dd, Polchinski:1999ry}.
In fact, one of the key ingredients of the AdS/CFT correspondence is the relation between the gravitational partition function in the bulk and the generating functional of correlation functions for the theory on the boundary~\cite{Gubser:1998bc,Witten:1998qj} which agrees with the extrapolate dictionary in AdS/CFT~\cite{Harlow:2011ke}.
Semi-classically, this implies that the gravitational path integral for a curved (asymptotically) AdS geometry localizes to the on-shell action contribution. This motivates us to look for the same principle for the S-matrix in AF spacetimes.\footnote{Here we focus on scalars and gauge fields minimally coupled to gravity, while for pure gravity this question has been addressed in the early 90s~\cite{Fabbrichesi:1993kz}; we are grateful to A.~Laddha for pointing out this work.} We will see that tree-level celestial correlators can indeed be viewed as being generated by the boundary on-shell action evaluated at null infinity, albeit with some caveats which we discuss.

Another way to utilize insights from AdS/CFT arises as follows. In a hyperbolic slicing of Minkowski space~\cite{deBoer:2003vf} the standard AdS/CFT dictionary can be applied on each slice and thus gives insight into celestial holography in flat space~\cite{Ball:2019atb,Cheung:2016iub}.
For certain curved AF geometries one would expect that, at least asymptotically, a similar slicing would exist. We will show that this is indeed the case for Schwarzschild and Kerr black holes, that is, there exists an asymptotic Euclidean AdS$_3$/CFT$_2$ structure. One might hope to exploit it to learn about the celestial CFT dual to bulk black holes in AF spacetimes, but we will leave this interesting question for future work.

The main focus of this work is to study celestial correlators for a class of non-trivial AF spacetimes, their infrared and ultraviolet properties, as well as their interpretation as standard CFT correlators.
We will consider backgrounds which admit a Kerr-Schild description, though our formalism will apply more generally whenever an S-matrix on the background exists.
Kerr-Schild geometries have many convenient features, for example they linearise the Einstein equations and they naturally provide examples of the classical double copy \cite{Monteiro:2014cda}.\footnote{See~\cite{stephani2009exact} for a discussion of Kerr-Schild geometries, while a recent review of the classical double copy can be found in~\cite{Kosower:2022yvp}.}
This provides a natural method to relate solutions of the Maxwell equations to those of Einstein gravity and we will consider scattering in both the electromagnetic and gravitational backgrounds. 

Our approach will be to first compute the two-point, or $1\to 1$, scattering amplitude of probe scalars in the leading Born approximation using the standard basis of momentum eigenstates\footnote{Going beyond the semiclassical approximation poses interesting but complicated challenges. For example, non-perturbative quantum effects at short distances like particle production can become relevant \cite{Hawking:1975vcx,Gibbons:1975kk,Gibbons:1975jb}, and it is not clear how a quantum S-matrix can be defined even for general AF backgrounds.}. Such amplitudes would be trivial in empty space but on curved geometries they encode information about the background solution and the couplings of the scalar field. We then recast these amplitudes in the basis of conformal boost eigenstates. 
Unlike their flat space counterparts, celestial two-point correlators on backgrounds exist at generic operator positions  and have other desirable features which we discuss. Perhaps most interestingly, in the case where the background itself has an interpretation of being generated by a conformal primary operator,
the celestial two-point correlators have exactly the form of CFT three-point functions computed in the standard vacuum with the non-trivial background encoded in the third operator. 

In addition to providing an arena for extending the flat space holographic description, scattering on non-trivial asymptotically flat backgrounds is of practical interest in the study of gravitational radiation. In this context, backgrounds which admit a point-particle description, such as some black hole solutions, are of particular interest. Particle-like backgrounds can be generated by classical three-point ``amplitudes'' with the off-shell coherent emission of one messenger particle - a photon or a graviton. They can be explicitly obtained via the classical limit of the in-in expectation value of the relevant quantum field in an on-shell state. This can be done through the KMOC formalism \cite{Kosower:2018adc,Monteiro:2020plf,Britto:2021pud} by using wavefunctions which are peaked around the value of the classical momentum in the $\hbar \to 0$ limit.\footnote{Henceforth we work in units where $\hbar=c=1$, keeping $G$ or $\kappa=\sqrt{32\pi G}$, and use $(-,+,+,+)$ signature.} Particle-like backgrounds we consider are the Coulomb potential of a point charge, its gravitational analogue given by the Schwarzschild metric and their ultraboost limits given by electromagnetic and gravitational shockwaves.

A related map between amplitudes and gravitational backgrounds occurs in the eikonal limit. In~\cite{tHooft:1987vrq} 't Hooft showed that the Aichelburg-Sexl shockwave solution \cite{Aichelburg:1970dh} can be used to compute the scattering of high-energy, gravitationally interacting scalar particles. In a frame where one of the particles is moving slowly the other can be viewed as generating the shockwave background and the $2\to2$ scattering amplitude can be computed semi-classically from the $1\to1$ amplitude in the non-trivial background. Generalisations of 't Hooft's result  were considered in \cite{Amati:1987uf, tHooft:1988oyr, Verlinde:1991iu,Kabat:1992tb}. In \cite{Adamo:2021rfq} it was shown that the semi-classical two-point amplitude for scalar fields on any stationary, linearized spacetime has the form of an eikonal amplitude and proposed a general relation which they tested for linearized Schwarzschild and Kerr geometries. Recently, an analogue of the eikonal limit for amplitudes in the celestial basis and its relation with celestial two-point functions in a shockwave background was considered \cite{deGioia:2022nkq}. To describe the scattering of massless particles with spin it is necessary to construct the spinning analogue of the Aichelburg-Sexl shockwave. A well known method for constructing a spinning solution  from a non-spinning solution, most notably the Kerr metric from Schwarzschild, is the Newman-Janis transformation \cite{Newman:1965tw}. An analogous transformation can be applied to the Aichelburg-Sexl shockwave \cite{Cristofoli:2020hnk} to produce what we will refer to as the spinning gravitational shockwave\footnote{The resulting metric is that found by \cite{Ferrari:1990tzs} as a result of boosting the Kerr metric along the direction of its spin. However its interpretation as due to a localised source is dubious at best. We are grateful to T.~Adamo and A.~Cristofoli for discussions on this point.}. One interesting feature of amplitudes on spinning geometries is that the Mellin transform has improved UV convergence properties which are possibly due to the finite-size effects of the spin.

Going beyond the leading order for the $1\to1$ amplitude and treating the background non-perturbatively requires exactly solving the wave equation. While the full result for general geometries is beyond our means, the computation can be organised as a sum over arbitrary interactions with the background. In the eikonal limit, where the momentum transferred to the probe is small and which captures the infrared divergent part of the amplitude, the expression can be resummed into an exponential. This exponentiation is familiar from the study of infrared divergences in scattering  amplitudes, long known in QED \cite{Yennie:1961ad} and gravity \cite{Weinberg:1965nx}, and provides a unique window into the non-perturbative low-energy dynamics of the theory. 
This allows one to address the issue of what are the correct asymptotic states in the presence of long-range interactions: the Faddeev-Kulish coherent state dressing \cite{Kulish:1970ut,Choi:2017ylo} guarantees the definition of a proper infrared finite S-matrix for such theories in four dimensions \cite{Kulish:1970ut, Ware:2013zja}. In addition, the simplicity of the result calls for an explanation in terms of (asymptotic) symmetries of the theory: remarkably, the celestial holography proposal makes this manifest \cite{Kapec:2017tkm,Choi:2017ylo,Himwich:2020rro,Arkani-Hamed:2020gyp}, and CCFT models for the soft dynamics have been proposed \cite{Cheung:2016iub,Nande:2017dba,Nguyen:2020hot, Nguyen:2021ydb,Kalyanapuram:2020epb,Pasterski:2021dqe,Kapec:2021eug}. 
We can then ask the same questions for AF backgrounds: can we understand the full CCFT description of the IR dynamics? How can we define infrared-finite correlators on backgrounds? In this work, we will address these questions for the simplest observable: the two-point function on point-like backgrounds.

This paper is organized as follows. In section~\ref{sec:scalarwavescattering} we study wave scattering on classical backgrounds. We focus on two-point amplitudes for massless scalars in Kerr-Schild backgrounds  that correspond to static and spinning sources both in gauge theory in section~\ref{ssec:KSgauge} and gravity in section~\ref{ssec:KSgravity}. We connect our results for two-point amplitudes on classical particle-like backgrounds to the eikonal limit of four-point tree-level amplitudes  in section~\ref{ssec:eikonal}. In section~\ref{sec:celestialwavescattering} we recast these results in CCFT. We compute the corresponding celestial two-point amplitudes for the various backgrounds in sections~\ref{sec:Celestial_static}-\ref{sec:Celestial_spinning} and discuss their soft limits in section~\ref{sec:Conformallysoft}. For shockwave backgrounds we show in section~\ref{sec:CelestialCorrelator} that the celestial amplitudes can be interpreted as three-point correlation functions between two massless asymptotic states and a conformal primary shockwave operator. In section~\ref{sec:bdyaction} we show that the boundary on-shell action localizes on the effective source in momentum-space and, similarly to AdS/CFT, gives the generating functional for the two-point amplitudes. Finally in section~\ref{sec:KSdressing}, we extend the conformal Faddeev-Kulish dressings of single particle states to dressings for particle-like backgrounds. In appendix~\ref{app:spinningwavescattering} and ~\ref{app:spinningbdyaction}, we extend the computation of amplitudes and the derivation of the on-shell action localization to the scattering of massless spinning fields on gauge and gravitational backgrounds. In appendix~\ref{app:slicing} we derive an asymptotic hyperbolic slicing representation for Schwarzschild and Kerr geometries, which provides some evidence for the presence of a conformal structure near the boundary of these AF geometries. 

\section{Wave scattering on backgrounds}\label{sec:scalarwavescattering}

To study wave scattering on classical backgrounds we use the method of Boulware and Brown~\cite{Boulware:1968zz}\footnote{See \cite{Monteiro:2011pc} for a nice review.}. This amounts to solving the classical equations of motion in the presence of a source. The  classical fields then serve as the generating functional of the tree-graph approximation to the corresponding quantum field theory.

Consider the generating functional $W[J]\equiv -i \log Z[J]$ for connected correlation functions defined by
\begin{equation}
 Z[J]=\int \mathcal D\Phi e^{i(S[\Phi]+J\Phi)},
\end{equation}
where $\Phi$ denotes the relevant dynamical field. The $n$th functional derivative of $W[J]$ with respect to $J$ yields the $n$-point correlator and amplitudes are computed via the LSZ procedure.
In the $\hbar \to 0$ limit the path integral is dominated by the classical solutions $\Phi_{cl}[J]$ 
of the equations of motion
and we find the relation
\begin{equation}
 \Phi_{cl}[J]=\frac{\delta W[J]}{\delta J}\,.
\end{equation}
The $n$-point connected correlator is thus obtained by differentiating the classical solution $n-1$ times with respect to the source.

We will be interested in scattering massless particles off various gauge and gravity backgrounds with a particle-like interpretation. The equations of motion are solved in momentum space order-by-order in perturbation theory 
\begin{equation}
 \bar \Phi_{cl}(p)=\sum_{n=0}^\infty \bar \Phi_{cl}^{(n)}(p),
\end{equation}
where $\bar \Phi_{cl}^{(n)}$ is of order $n$ in the coupling and we use the conventions for the Fourier transform
\begin{equation}
    \Phi_{cl}(x)=\int \frac{d^4p}{(2\pi)^4} e^{ip\cdot x}\bar \Phi_{cl}(p)\,, \quad \bar \Phi_{cl}(p)=\int d^4x e^{-ip\cdot x} \Phi_{cl}(x)\,.
\end{equation}
The $n$-point amplitude is given by
\begin{equation}
\label{eq:npointA}
 \mathcal A_n(p_1,\dots, p_n)=i^n \prod_{k=1}^n (\lim_{p_k^2\to 0}p_k^2) \left. \frac{\delta }{\delta J(p_{1})}\dots\frac{\delta}{\delta J(p_{n-1})}\bar \Phi_{cl}(-p_n)\right|_{J=0}.
\end{equation}
In the following we will focus on two-point amplitudes of massless particles in backgrounds of Kerr-Schild type. In gravity these take the form
\begin{equation}
   g_{\mu\nu}(x)=\eta_{\mu\nu}+h_{\mu\nu}(x), \quad h_{\mu\nu}(x)=k_\mu k_\nu V(x), 
\end{equation}
while in gauge theory
\begin{equation}
\label{eq:pot_KS}
 A_\mu(x)=k_\mu V(x).
\end{equation}
The scalar function $V$ satisfies the free wave equation $\eta^{\mu\nu}\partial_\mu \partial_\nu V(x)=0$ and the Kerr-Schild vector $k^\mu$ is null and geodesic
\begin{equation}
    \eta_{\mu\nu}k^\mu k^\nu =0=g_{\mu\nu}k^\mu k^\nu, \quad k^\mu \partial_\mu k_\nu=0=k^\mu \nabla_\mu k_\nu.
\end{equation}
Backgrounds that are of Kerr-Schild type include the Schwarzschild and Kerr black holes as well as their ultraboosted limits, the Aichelburg-Sexl and spinning shockwave metrics. They can be obtained via the classical double copy of the gauge theory backgrounds corresponding to the Coulomb field of static and spinning point charges and their ultraboosted limits.

We now derive the two-point amplitudes on gauge and gravitational Kerr-Schild backgrounds focusing on scattering minimally coupled complex massless scalars, while the generalization to spinning probes is detailed in appendix~\ref{app:spinningwavescattering}. 

\subsection{Scattering on Kerr-Schild gauge theory backgrounds}\label{ssec:KSgauge}

The wave equation for a complex scalar field $\phi(x)$ in scalar QED in the presence of a source $J(x)$ is
\begin{equation}\label{eq:gaugebgd}
 \eta^{\mu\nu} D_\mu D_\nu \phi(x)=J(x),
\end{equation}
with the gauge covariant derivative $D_\mu =\partial_\mu-ieA_\mu$.
To solve 
\begin{equation}
\label{eq:em_we}
    \eta^{\mu\nu}\partial_\mu \partial_\nu \phi-2i e A^{\mu}\partial_\mu  \phi-i e\partial_\mu A^{\mu} \phi-e^2 A_\mu A^\mu\phi=J
\end{equation}
we recast it in terms of a wave equation with an effective source
\begin{equation}
\label{eq:em_Jeff}
    \eta^{\mu\nu}\partial_\mu \partial_\nu \phi = J_{\text{eff}}\,, \qquad J_{\text{eff}} \equiv J + 2i e A^{\mu}\partial_\mu  \phi + i e\partial_\mu A^{\mu} \phi + e^2 A_\mu A^\mu\phi \,.
\end{equation}
Then we go to Fourier space and we work perturbatively in the coupling $e$
\begin{equation}
\label{eq:em_mom_we}
 -p^2 \bar \phi(p) =\bar J(p) - e\int \frac{d^4p'}{(2\pi)^4} \bar A^\mu(p-p') (p_\mu+p'_\mu)  \bar \phi(p')+\mathcal O(e^2).
\end{equation}
At leading order this is solved by $\bar \phi^{(0)}(p)=-\frac{\bar J(p)}{p^2}$ while at subleading order
\begin{equation}
\bar \phi^{(1)}(p)=\frac{e}{p^2} \int\frac{d^4p'}{(2\pi)^4}\bar   A^\mu(p-p')  (p_\mu+p'_\mu)\bar \phi^{(0)}(p').
\end{equation}
The two-point amplitude for real potentials, up to order $\mathcal O(e)$ is
\begin{equation}
\badat{2}
 \mathcal A_2(p_1,p_2)&=-\lim_{p_1^2\to 0}\lim_{p_2^2\to 0} p_1^2p_2^2 \frac{\delta\bar \phi(-p_2)}{\delta \bar J(p_1)}\\
 &=(2\pi)^4  \lim_{p_1^2\to0}\lim_{p_2^2\to0} p_1^2 \delta^{(4)}(p_1+p_2)+e(p_1-p_2)_\mu \bar A^\mu(p_1+p_2).
\eadat
\end{equation}
In the following we compute
\begin{equation}
 \mathcal A_2^{(1)}(p_1,p_2)= e(p_1-p_2)_\mu \bar A^\mu(p_1+p_2)
\end{equation}
for various Kerr-Schild backgrounds.

\subsubsection{Coulomb}
%
We consider the Coulomb field of a static point charge
\begin{equation}
 A_\mu(x)=V(x)k_\mu\,,\quad V(x)=\frac{Q}{4\pi r}
\end{equation}
with Kerr-Schild vector
\begin{equation}
    k_\mu=u_\mu-n_\mu\,, \quad u_\mu=(-1,0,0,0)\,, \quad n_\mu=\partial_\mu r\,.
\end{equation}
Noting that $n_\mu=r\partial_\mu \log r$ we can express the gauge potential as 
\begin{equation}
    A_\mu(x)=\frac{Q}{4\pi r} u_\mu+\partial_\mu \Lambda\,,  
\end{equation} 
where $\Lambda=\frac{Q}{4\pi} \log r$, and henceforth drop the total derivative term since it does not contribute to the amplitude.
The gauge potential in momentum-space is given by
\begin{equation}
\bar A_\mu(p)=\frac{2\pi Q}{p^2} \delta(p\cdot u)u_\mu,
\end{equation}
where we used
\begin{equation}\label{eq:coulombpotential}
 \int d^3 \vec{x} \frac{e^{i\vec{p}\cdot \vec{x}}}{r}=\frac{4\pi}{|\vec{p}|^2},
\end{equation}
regulating the integral by $e^{-\mu r}$ as $\mu \to 0$.
This yields the two-point amplitude
\begin{equation}\label{eq:2ptCoulomb}
    \mathcal A_2^{(1)}(p_1,p_2)=4\pi eQ \frac{(p_1\cdot u)\delta((p_1+p_2)\cdot u)}{(p_1+p_2)^2}.
\end{equation}

\subsubsection{$\sqrt{\text{Kerr}}$}
The gauge field corresponding to a spinning charged particle, and which is the single copy of the Kerr solutions \cite{Monteiro:2014cda}, is given by\footnote{Here we follow the rewriting of the Kerr metric in \cite{Vines:2017hyw}.}
\begin{equation}\label{eq:KS_kerr_vec}
    A_\mu(x)= V(x) k_\mu\,, \quad V(x)= \frac{Q}{4\pi}\frac{\tilde r}{\Sigma}\,,
\end{equation}
with Kerr-Schild vector
\begin{equation}
    k_\mu=u_\mu-\tilde n_\mu\,, \quad u_\mu=(-1,0,0,0)\,, \quad \tilde n_\mu= \frac{\Sigma}{2\tilde r} \partial_\mu \log(\tilde r^2+a^2)+u^\lambda \epsilon_{\lambda \mu \rho \sigma}\frac{r^\rho a^\sigma}{\tilde r^2+a^2}\,.
\end{equation}
In terms of oblate spheroidal coordinates $(\tilde r, \tilde \theta, \phi)$ defined by $x+iy=\sqrt{\tilde r^2+a^2}\sin \tilde \theta e^{i\phi}$ and $z=\tilde r \cos \tilde \theta$ we have $\Sigma=\tilde r^2+a^2 \cos^2\tilde \theta$ and $\tilde r$ is defined through 
\begin{equation}
    \frac{x^2+y^2}{\tilde{r}^2+a^2}+\frac{z^2}{\tilde{r}^2}=1~.
\end{equation}
Furthermore, we have $r_\mu =\frac{1}{2}\partial_\mu r^2$ and we take the covariant spin vector $a^\mu=(0,0,0,a)$.
The gauge potential can then be expressed as 
\begin{equation}
A_\mu(x)=\frac{Q}{4\pi}\frac{\tilde r}{\Sigma}\left(u_\mu-u^\lambda \epsilon_{\lambda\mu \rho \sigma}\frac{r^\rho a^\sigma}{\tilde r^2+a^2}\right)+\partial_\mu \Lambda\,,   
\end{equation}
where $\Lambda=\frac{Q}{8\pi}\log(\tilde r^2+a^2)$.

To compute the Fourier transformed potential we again drop the pure gauge term and make use of the identities $\frac{\tilde r}{\Sigma}=\cos(\vec{a}\cdot \vec{\nabla})\frac{1}{r}$ and $\frac{\tilde r}{\Sigma}\frac{\vec{r}\times \vec{a}}{\tilde r^2+a^2}=\vec{a}\times \vec{\nabla}\frac{\sin(\vec{a}\cdot \vec{\nabla})}{(\vec{a}\cdot \vec{\nabla})}\frac{1}{r}$. This yields
\begin{equation}
    \bar A_\mu(p)=\frac{2 \pi Q}{p^2} \delta(p\cdot u)\Big[u_\mu \cosh(a\cdot p) +i u^\nu \epsilon_{\nu\mu\alpha\beta}a^\alpha p^\beta\frac{\sinh (a\cdot p)}{(a\cdot p)}\Big]\,.
\end{equation}
The leading non-trivial term in the two-point amplitude is
   \begin{equation}
   \badat{2}
    \mathcal A_2^{(1)}(p_1,p_2)&=4\pi e Q \frac{\delta((p_1+p_2)\cdot u)}{(p_1+p_2)^2}\Big[(p_1\cdot u) \cosh (a\cdot (p_1+p_2))\\
 &\kern+120pt -i  \epsilon_{\mu \nu \alpha \beta}u^\mu a^\nu p_1^\alpha p_2^\beta \frac{\sinh(a\cdot(p_1+p_2))}{a\cdot (p_1+p_2))}\Big]~.
 \eadat
\end{equation}

\subsubsection{Electromagnetic shockwave}

The gauge field corresponding to an ultra-boosted point charge is given by 
\begin{equation}
    A_\mu(x)=V(x) k_\mu\,, \quad V(x)=-\frac{Q}{4\pi} \log(x^2) \delta(k\cdot x)\,.
\end{equation}
Without loss of generality we take $k_\mu=u_\mu=(-1,0,0,1)$ so that we can write $\log(x^2)\delta(k\cdot x)=\log\left( (x^1)^2+(x^2)^2\right) \delta(x^0-x^3)$.
In momentum-space the gauge potential becomes
\begin{equation}
   \bar A_\mu(p)=\frac{2\pi Q}{p^2}\delta(p^0-p^3)u_\mu.
\end{equation}
To obtain this expression we changed variables $(x^1,x^2)=\rho (\cos \varphi, \sin \varphi)$ and used 
\begin{equation}\label{eq:logJ0nonspinning}
\int_0^\infty d\rho \rho \log \rho J_0(\rho \sqrt{(p^1)^2+(p^2)^2})=\frac{-1}{(p^1)^2+(p^2)^2}
\end{equation}
Note that the $\rho$ integral can be solved from the integral over the product of (modified) Bessel functions
\begin{equation}\label{eq:K0J0nonspinning}
    \int_0^\infty d\rho \rho K_0(b\rho)J_0(c \rho)=\frac{1}{b^2+c^2}, \quad {\rm Re}(b)>0,\; {\rm Re}(c)>0,\; {\rm Im}(b)=0,\; {\rm Im}(c)\leq 0
\end{equation}
in an expansion near $0<|z|\ll 1$ where $K_0(z)\sim -\log \frac{z}{2}-\gamma_E$ with $\gamma_E$ denoting the Euler-Mascheroni constant. From $b=2\mu e^{-\gamma_E}$ with $\mu\ll 1$ the leading $\mathcal O(\mu^0)$ term in~\eqref{eq:K0J0nonspinning} is $\frac{-1}{c^2}$ which equates to~\eqref{eq:logJ0nonspinning}. 
A generalization of this argument will be useful for the spinning case below. The two-point amplitude in the electromagnetic shockwave background  is
\begin{equation}\label{eq:2ptEMshock}
     \mathcal A_2^{(1)}(p_1,p_2)=-4\pi e Q  \frac{p_1^-\delta(p_1^-+p_2^-)}{(p_1+p_2)^2}\,,
\end{equation}
where we used null coordinates $p^\pm=\frac{1}{2}(p^0\pm p^3)$.

\subsubsection{Spinning electromagnetic shockwave}
Ultra-boosting the $\sqrt{\text{Kerr}}$ solution along its axis of rotation produces a spinning shockwave geometry that is of the Kerr-Schild form \cite{Ferrari:1990tzs}. The single copy produces an electromagnetic spinning shockwave and is given by
\begin{equation}
    A_\mu(x)=V(x) k_\mu\,, \quad V(x)=-\frac{Q}{4\pi} \log (x^2-a^2) \delta(k\cdot x)\,,
\end{equation}
where $a$ is the spin parameter and again we take $k_\mu=u_\mu=(-1,0,0,1)$. A slightly more involved calculation than in the non-spinning case gives
\begin{equation}
     \bar A_\mu(p)=\pi^2 Q \frac{ i a}{|p|}H_{-1}^{(2)}(a|p|)\delta(p^0-p^3)u_\mu,
\end{equation}
in terms of a Hankel function where $|p|\equiv \sqrt{p^2}$.
To arrive at this result we follow the same arguments as above changing to polar coordinates, replacing the logarithm by the modified Bessel function, computing the integral
\begin{equation}\label{eq:K0J0spinning}
    \int_0^\infty d\rho \rho K_0(b\sqrt{\rho^2-a^2})J_0(c \rho)(\rho^2-a^2)^{-\mu/2}=\frac{\pi}{2} e^{i\pi/2} \frac{a}{\sqrt{b^2+c^2}}H_{-1}^{(2)}(a\sqrt{b^2+c^2}),
\end{equation}
for ${\rm Re}(\mu)<1$, $a>0$, $b>0$, $c>0$, and expanding the result with $b=2\mu e^{-\gamma_E}$ up to $\mathcal O(\mu^0)$.
The two-point amplitude in the spinning electromagnetic shockwave background is
\begin{equation}\label{eq:2ptspinningEMshock}
    \mathcal A_2^{(1)}(p_1,p_2)=-2\pi^2 e Q i a H_{-1}^{(2)}(a|p_1+p_2|) \frac{p_1^-\delta(p_1^-+p_2^-)}{|p_1+p_2|} .
\end{equation}
Note that $\frac{i\pi}{2}\frac{a}{|p|} H_{-1}^{(2)}(a|p|) \to \frac{1}{p^2}$ as $a\to 0$. Thus in the limit $a\to 0$ we recover the non-spinning result~\eqref{eq:2ptEMshock}.

\subsection{Scattering on Kerr-Schild gravity backgrounds}\label{ssec:KSgravity}

The wave equation for a complex scalar field $\phi(x)$ minimally coupled to gravity and in the presence of a source $J(x)$ is\footnote{In principle we could include additional couplings, for example to the scalar curvature, $\sqrt{-g}\mathcal{R}\phi^2$, but we set these to zero for simplicity.}
\begin{equation}
  \frac{1}{\sqrt{-g}}\partial_\mu(\sqrt{-g}g^{\mu\nu} \partial_\nu \phi(x))=J(x)~.
\end{equation}
Quite generally, we can study this equation perturbatively about the flat background where the metric is $g_{\mu\nu}=\eta_{\mu \nu}+h_{\mu\nu}$ with $h_{\mu \nu}$ taken to be small. 
The wave equation can be written in terms of an effective source,
\begin{equation}
\label{eq:grav_scal_Jeff}
    \eta^{\mu\nu}\partial_\mu \partial_\nu \phi = J_{\text{eff}} \,, \qquad J_{\text{eff}} \equiv J +h^{\mu\nu}\partial_\mu \partial_\nu \phi+\partial_\mu h^{\mu\nu} \partial_\nu \phi-g^{\mu \nu}\partial_\mu \log (\sqrt{-g}) \partial_\nu \phi \,.
\end{equation}
The wave equation becomes, to linear order in the metric perturbation,
\begin{equation}
\label{eq:grav_scal_weq}
    \eta^{\mu\nu}\partial_\mu \partial_\nu \phi=J+h^{\mu\nu}\partial_\mu \partial_\nu \phi+\partial_\mu (h^{\mu\nu}-\tfrac{1}{2} h^\lambda{}_\lambda \eta^{\mu \nu})\partial_\nu \phi
\end{equation}
where all indices are raised using the flat metric. We again solve this equation by going to Fourier space and working perturbatively in the background perturbation. The equation of motion can be written as
\begin{equation}
 -p^2 \bar \phi(p)=\bar J(p)-\int \frac{d^4p'}{(2\pi)^4} \bar h^{\mu\nu}(p-p') \big[p_\mu p'_\nu-\tfrac{1}{2}\eta_{\mu \nu} (p\cdot p'-p'^2)\big] \bar \phi(p'),
\end{equation}
so that
at zeroth order we have $\bar \phi^{(0)}(p)=-\frac{\bar J(p)}{p^2}$ while at linear order we get
\begin{equation}
\bar \phi^{(1)}(p)=\frac{1}{p^2} \int\frac{d^4p'}{(2\pi)^4}\bar   h^{\mu\nu}(p-p')  \big[p_\mu p'_\nu -\tfrac{1}{2}\eta_{\mu \nu}(p\cdot p'-p'^2)\big]\bar \phi^{(0)}(p').
\end{equation}
The two-point amplitude up to linear order in the perturbation is
\begin{equation}
\badat{3}
 \mathcal M_2(p_1,p_2)&=-\lim_{p_1^2\to 0}\lim_{p_2^2\to 0} p_1^2p_2^2 \frac{\delta\bar \phi(-p_2)}{\delta \bar J(p_1)}\\
 &=(2\pi)^4  \lim_{p_1^2\to0}\lim_{p_2^2\to0} p_1^2 \delta^{(4)}(p_1+p_2)\\
 &  \quad -\big[(p_1)_\mu (p_2)_\nu -\tfrac{1}{2} \eta_{\mu \nu} (p_1\cdot p_2)\big]\bar h^{\mu\nu}(p_1+p_2)~.
\eadat
\end{equation}
The amplitude
\begin{equation}
\label{eq:grav_2pt}
 \mathcal M_2^{(1)}(p_1,p_2)=-\big[(p_1)_\mu (p_2)_\nu -\tfrac{1}{2} \eta_{\mu \nu} (p_1\cdot p_2)\big] \bar h^{\mu\nu}(p_1+p_2)
\end{equation}
is invariant under linearized diffeomorphisms of the background, $\bar{h}_{\mu \nu}(p)\to \bar{h}_{\mu\nu}(p)+p_{(\mu} \xi_{\nu)}$, as can be seen by substitution and use of the on-shell momentum conditions. 
We will be interested in computing \eqref{eq:grav_2pt} in various Kerr-Schild backgrounds for which
\begin{equation}
    h_{\mu\nu}(x)=k_\mu k_\nu V(x), 
\end{equation} 
which implies that $g^{\mu\nu}=\eta^{\mu\nu}-k^\mu k^\nu V$ and $\sqrt{-g}=1$ so that the wave equations simplifies and \eqref{eq:grav_scal_weq} is in fact exact. Nonetheless,  we still take the perturbation to be small in solving for the scalar field and the two-point function is given by \eqref{eq:grav_2pt}, though with the trace term vanishing due to $k^2=0$. 
\subsubsection{Schwarzschild}
We first consider scattering in the Schwarzschild geometry which in Kerr-Schild form is given by
\begin{equation}
    h_{\mu\nu}(x)=V(x)k_\mu k_\nu \,, \quad V(x)=\frac{2G M }{r}\,.
\end{equation}
The Kerr-Schild vector 
\begin{equation}
    k_\mu=u_\mu-n_\mu\,, \quad u_\mu=(-1,0,0,0)\,, \quad n_\mu=\partial_\mu r
\end{equation}
is the same as in the Coulomb case. This allows us to write~\cite{Vines:2017hyw}\footnote{Our conventions are $A_{(a}B_{b)}=\frac{1}{2}(A_a B_b+A_b B_a)$. }
\begin{equation}
    h_{\mu\nu}(x)=\frac{4GM}{r}\left(u_\mu u_\nu -\frac{1}{2}u^2\eta_{\mu\nu}\right)+2\partial_{(\mu}\xi_{\nu)}\,, 
\end{equation}
where $\xi^\mu=-GM\left[\log (r^2) u^\mu+n^\mu\right]$ and we can drop the pure diffeomorphism term henceforth.
The Fourier transformed gravitational potential is
\begin{equation}
      \bar h_{\mu\nu}(p)=\frac{32\pi^2 GM}{p^2} \delta(p\cdot u) \left(u_\mu u_\nu -\frac{1}{2}u^2\eta_{\mu\nu}\right)\,,
\end{equation}
where we used again~\eqref{eq:coulombpotential}. Thus the two-point amplitude is
\begin{equation}
     \mathcal M^{(1)}_2(p_1,p_2)=32\pi^2 GM (p_1\cdot u)^2 \frac{\delta((p_1+p_2)\cdot u)}{(p_1+p_2)^2}~.
\end{equation}

\subsubsection{Kerr}
The Kerr geometry can also be written in Kerr-Schild form
\begin{equation}
    h_{\mu\nu}= V(x) k_\mu k_\nu\,, \quad V(x)=2 GM \frac{\tilde r}{\Sigma}\,,
\end{equation}
with the same Kerr-Schild vector
\begin{equation}
    k_\mu=u_\mu-\tilde n_\mu\,, \quad u_\mu=(-1,0,0,0)\,, \quad \tilde n_\mu= \frac{\Sigma}{2\tilde r} \partial_\mu \log(\tilde r^2+a^2)+u^\lambda \epsilon_{\lambda \mu \rho \sigma}\frac{r^\rho a^\sigma}{\tilde r^2+a^2}\,,
\end{equation}
as for its electromagnetic analogue. Again we take  $a^\mu=(0,0,0,a)$.
One can then show~\cite{Vines:2017hyw} that the gravitational potential can be expressed as
\begin{equation}
\label{eq:h_kerr}
    h_{\mu\nu}(x)=4GM\frac{\tilde r}{\Sigma} \left(u_\mu u_\nu -\frac{1}{2}u^2\eta_{\mu\nu} +u_{(\mu}\epsilon_{\nu)\lambda \rho \sigma} u^\lambda \frac{r^\rho a^\sigma}{\tilde r^2+a^2}\right)+2\partial_{(\mu} \xi_{\nu)}\,, 
\end{equation}
where $\xi^\mu=-GM\left[\log(\tilde r^2+a^2)u^\mu+\tilde n^\mu\right]$, and we can drop the linearised diffeomorphism since it will drop out of the amplitude at the order to which we work. We use the same identities as in the electromagnetic case to obtain the Fourier transformed gravitational potential
\begin{equation}
    {\bar{h}}_{\mu\nu}(p)=\frac{32\pi^2 GM }{p^2}\delta(u\cdot p)\Big[\left(u_\mu u_\nu-\frac{1}{2}u^2\eta_{\mu\nu}\right) \cosh(a\cdot p)-i u_{(\mu}\epsilon_{\nu)\lambda\rho\sigma} u^\lambda a^\rho p^\sigma \frac{\sinh(a\cdot p)}{a\cdot p}\Big]\,.
\end{equation}
The two-point amplitude is thus found to be 
\begin{equation}
\badat{2}
    \mathcal M^{(1)}_2(p_1,p_2)&=32\pi^2 GM \frac{\delta( (p_1+p_2)\cdot u)}{(p_1+p_2)^2}(p_1\cdot u) \Big[(p_1\cdot u) \cosh(a\cdot (p_1+p_2))\\
    &\kern+140pt   -i  \epsilon_{\mu \nu \alpha \beta}u^\mu a^\nu p_1^\alpha p_2^\beta \frac{\sinh(a\cdot (p_1+p_2))}{a\cdot (p_1+p_2)}\Big]\,.
\eadat
\end{equation}

\subsubsection{Gravitational shockwave}
The Aichelburg-Sexl geometry describing the gravitational field of an ultra-relativistic black hole or a massless particle is 
\begin{equation}
    h_{\mu\nu}=V(x)k_\mu k_\nu \,, \quad V(x)=-4 G P^+\log(x^2)\delta(k\cdot x) \,,
\end{equation}
where we take $k_\mu=u_\mu=(-1,0,0,1)$ and $P^+$ is the lightcone energy of the shockwave. Following the same steps as in the gauge theory case we find
\begin{equation}
   \bar h_{\mu\nu}(p)=\frac{32 \pi^2G P^+  }{p^2}\delta(p^0-p^3)u_\mu u_\nu
\end{equation}
so that the two-point amplitude is 
\begin{equation}\label{eq:2ptgravityshock}
  \mathcal M_2^{(1)}(p_1,p_2)=-64 \pi^2GP^+ \frac{p_1^-p_2^-\delta(p_1^-+p_2^-)}{(p_1+p_2)^2}.
\end{equation}

\subsubsection{Spinning gravitational shockwave}
As previously described, ultra-boosting the $\sqrt{\text{Kerr}}$ solution along its axis of rotation produces a spinning shockwave geometry \cite{Ferrari:1990tzs}. This geometry shares some features, though it is different, than gyraton metric describing the spinning analog of a highly boosted spinning particle or beam of radiation with angular momentum \cite{Frolov:2005in, Frolov:2005zq}. As we will later see, for our purposes it is the fact that it corresponds to a conformal primary operator that is most important. The spinning shockwave metric is 
\begin{equation}
    h_{\mu\nu}(x)= V(x)k_\mu k_\nu\,, \quad V(x)=-4 G P^+ \log(x^2-a^2) \delta(k\cdot x)\,,
\end{equation}
and we take again $k_\mu=u_\mu(-1,0,0,1)$.
In momentum-space the gravitational potential becomes
\begin{equation}
      \bar h_{\mu\nu}(p)=16 G P^+\pi^3 \frac{ ia}{|p|}H_{-1}^{(2)}(a|p|)\delta(p^0-p^3)u_\mu u_\nu.
\end{equation}
The two-point amplitude is
\begin{equation}\label{eq:2ptspinninggravityshock}
    \mathcal M_2^{(1)}(p_1,p_2)=-32\pi^3 GP^+ i a H_{-1}^{(2)}(a|p_1+p_2|) \frac{p_1^-p_2^-\delta(p_1^-+p_2^-)}{|p_1+p_2|} ,
\end{equation}
where for $a\to 0$ we recover the non-spinning result~\eqref{eq:2ptgravityshock}.

\subsection{Particle-like backgrounds and the classical amplitude matching}\label{ssec:eikonal}
\label{sec:eikonal}

\begin{figure}[!htb]
\centering
\includegraphics[scale=0.9]{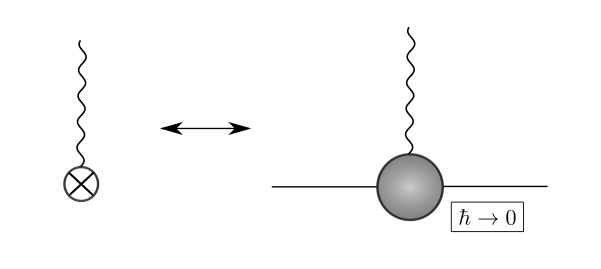}
\caption{Equivalence between point-particle backgrounds and the 3-pt function with an off-shell coherent emission.}
\label{fig:3point}
\end{figure}

Particle-like backgrounds arise from considering some kinematic limit of a Lorentz-invariant field theory particle description in flat space. They can be generated by classical three-point ``amplitudes'' with the off-shell coherent emission of one messenger particle,  a photon or a graviton as figuratively shown in Fig.\ref{fig:3point}.

We can define particle-like backgrounds from the in-in expectation value
\begin{align}
\Psi_{I}^{\text{particle-like}}(x) \equiv & \langle \psi |S^{\dagger} \hat{\Psi}_{I}(x) S | \psi \rangle \Big|_{\hbar \to 0} \,, \qquad \hat{\Psi}_{I} = \hat{\phi}, \hat{A}_{\mu},\hat{h}_{\mu \nu} \,,
\end{align}
where $S$ is the flat space S-matrix in the theory, $\hat{\Psi}_{I}(x)$ represents the field operator which contains both on-shell and off-shell contributions in Lorentzian signature and 
\begin{align}
| \psi \rangle &= \int d \Phi (p) \psi_{u}(p) \ket{p} 
\end{align}
is the on-shell external particle state smeared with an appropriate wavefunction $\psi_{u}(p)$ peaked along the classical trajectory of 4-velocity $u^{\mu}$ \cite{Kosower:2018adc,Monteiro:2020plf}. An equivalent procedure consists in matching amplitudes with effective classical sources in the $\hbar \to 0$ limit \cite{Bjerrum-Bohr:2018xdl}.

\begin{figure}[!htb]
\centering
\includegraphics[scale=0.9]{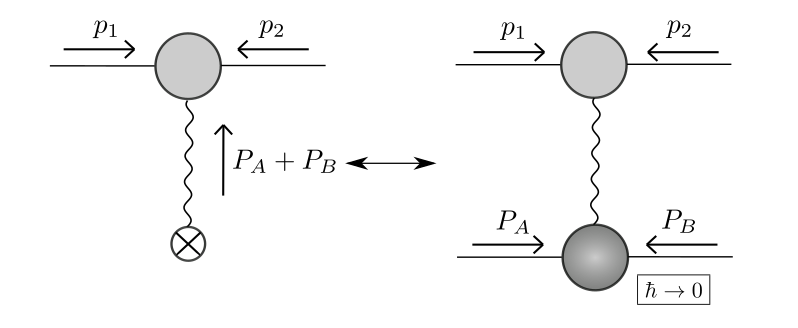}
\caption{Equivalence between wave scattering on point-particle backgrounds and the four-point function, where the background is effectively identified with the classical limit of the dark gray interaction vertex of the four-point amplitude. }
\label{fig:4point}
\end{figure}

When considering wave scattering on classical point-like backgrounds, we can construct a map from the four-point amplitude\footnote{In general, the scattering of a wave on a background may require considering higher-point amplitudes but we are interested here in the leading connected amplitude contribution. For classical wave scattering, this can be made explicit by using coherent states \cite{Cristofoli:2021vyo} and in such case only the four-point amplitude contributes at all orders. } to the two-point function on the background. This is illustrated in figure~\ref{fig:4point}. 
At leading order in the coupling, this procedure is entirely equivalent to the matching with the eikonal four-point tree-level amplitude described in a recent work \cite{Adamo:2021rfq}, and indeed we will check explicitly in the following the consistency of such an approach with our solutions of the wave equation in the leading Born approximation. Nevertheless, our prescription would differ at higher orders in perturbation theory because in holography we want to also include the quantum terms in the wave scattering.\footnote{See \cite{Bjerrum-Bohr:2014zsa} for an explicit calculation of these type of (long-range) quantum contributions at one loop.}

We illustrate the matching between eikonal four-point amplitudes in gauge theory and gravity and two-point amplitudes on classical backgrounds for the non-spinning examples considered in the previous section.
The Coulomb and electromagnetic shockwave solutions are both special cases of potentials sourced by point-like charge, with trajectory $y^\mu(\tau)$ and four-velocity $u^\mu=dy^\mu/d\tau$ corresponding to a current $j^\mu(x)=-Q\int d\tau~ u^\mu(\tau) \delta^{(4)}(x- y)$.
 The resulting  two-point amplitude is
\begin{equation}
\mathcal{A}_2^{(1)}(p_1,p_2) = e Q  \int d\tau ~\frac{(p_1-p_2)\cdot u}{(p_1+p_2)^2} e^{-i  (p_1+p_2)\cdot y}
\end{equation}
which for uniform motion, $y^{\mu}=\tau u^\mu$ with $u^\mu$ constant, gives
\begin{equation}
\label{eq:2ptemgen}
\mathcal{A}_2^{(1)}(p_1,p_2) =2\pi e Q  \frac{(p_1-p_2)\cdot u}{(p_1+p_2)^2} \delta((p_1+p_2)\cdot u)~
\end{equation}
This describes both the Coulomb, $u^\mu=(1,0,0,0)$, and electromagnetic shockwave, $u^\mu=(1,0,0,1)$, cases. 
Similarly, we can view the Schwarzschild and Aichelburg-Sexl  geometries as being sourced by point-like charges. The harmonic-gauge, linearized Einstein equations can be written as 
\begin{equation}
\label{eq:lin_hg_eeq}
\Box h_{\mu \nu}=-\frac{\kappa^2}{2} \mathcal{P}_{\mu\nu\alpha \beta} T^{\alpha\beta}
\end{equation}
where we have introduced the perturbative coupling $\kappa=\sqrt{32\pi G}$ and $\mathcal{P}^{\mu\nu}{}_{\alpha \beta}=\delta_{(\mu}{}^{(\alpha}\delta_{\nu)}{}^{\beta)}-1/2 \eta_{\mu \nu}\eta^{\alpha\beta}$ is the flat space trace-reverser. The stress-tensor can be written to this order, \cite{Hanson:1974qy, Bailey:1975fe}\footnote{See also the MPD approach to extended bodies in General Relativity \cite{Mathisson:1937zz, papapetrou1951spinning, tulczyjew1959motion, dixon1970dynamics, dixon2015new}},
in an effective, skeleton, form with support along a world-line $y^\mu(\tau)$
\begin{equation}
T^{\mu \nu}=\int d\tau ~ \hat{\mathcal{T}}^{\mu \nu} \delta^{(4)}(x-y)
\end{equation}
with  $\hat{\mathcal{T}}^{\mu\nu}$ a differential operator given in terms of the moments of $T^{\mu\nu}$ and depending on fields defined along the particle world-line. For the case of uniform, spinless motion $\hat{\mathcal{T}}^{\mu \nu}=r_0 u^\mu u^\nu$, with $r_0$ and $u^\mu$ constant, the two-point amplitude is
\begin{equation}
\label{eq:2ptgravgen}
    \mathcal M_2^{(1)}(p_1,p_2)=-32\pi^2 G {r}_0 \frac{(p_1\cdot q)( p_2\cdot q) \delta((p_1+p_2)\cdot u)}{(p_1+p_2)^2}.
\end{equation}
The Schwarzschild geometry corresponds to $u^\mu=(1,0,0,0)$, $r_0=M$ and  the Aichelburg-Sexl background to $u^\mu=(1,0,0,1)$ with $r_0=P^+$ the shockwave light-cone energy.

Following the derivation in \cite{Saotome:2012vy,Adamo:2021rfq}, at leading order in the coupling we can identify point-particle backgrounds with a particular choice of the kinematics for the point-particle entering in the four-point eikonal amplitude.
We can therefore match the tree-level two-point function in the Born approximation with the tree-level classical eikonal amplitude where the momentum-conserving delta function is stripped off.
For the Coulomb and Schwarzschild backgrounds, corresponding to $u^\mu=(1,0,0,0)$, the two-point amplitudes of massless scalars with momenta $p_1,p_2$ can be expressed in terms of the stripped eikonal four-point amplitudes 
\begin{equation} \label{2pt4ptmatch}
    \badat{2}
\mathcal{A}^{(1)}_{2}(p_1,p_2) & = \frac{2 \pi \delta((p_1+ p_2)\cdot u)}{4 M} \mathcal{A}_{4,\text{EM}}^{(1),\text{eik}}(s,t) \Big|_{P_A^{\mu} = P_B^{\mu} = P^\mu} \,, \\ 
\mathcal{M}^{(1)}_{2}(p_1,p_2)  &= \frac{2 \pi \delta((p_1 + p_2)\cdot u)}{4 M} \mathcal{M}_{4,\text{GR}}^{(1),\text{eik}}(s,t) \Big|_{P_A^{\mu} = P_B^{\mu} = P^\mu},
 \eadat
\end{equation} 
where the momenta $P^\mu= M u^{\mu}$ of the massive scalars are identified with the background. The matching \eqref{2pt4ptmatch} is achieved in the eikonal limit $-t\ll s$,
\begin{equation}
    s=-(p_1+P)^2\,, \quad t=-(p_1+p_2)^2\,,
\end{equation} 
which corresponds exactly to the classical limit for the leading order wave scattering on a point-like background. The amplitudes involving the electromagnetic and gravitational shockwave backgrounds, corresponding to $u^\mu=(1,0,0,1)$, become
\begin{equation}
    \badat{2}
\mathcal{A}^{(1)}_{2}(p_1,p_2) & = \frac{2 \pi \delta((p_1+ p_2)\cdot u)}{2 P_A^{+}} \mathcal{A}_{4,\text{EM}}^{(1),\text{eik}}(s,t) \Big|_{P_A^{\mu} = P_B^{\mu} = P^\mu} \,, \\ 
\mathcal{M}^{(1)}_{2}(p_1,p_2)  &= \frac{2 \pi \delta((p_1 + p_2)\cdot u)}{2 P_A^{+}} \mathcal{M}_{4,\text{GR}}^{(1),\text{eik}}(s,t) \Big|_{P_A^{\mu} = P_B^{\mu} = P^\mu}   \,.
 \eadat
\end{equation}
where in this case $P^\mu=P_A^{+} q^{\mu}$ with $P_A^{+} =\tfrac{1}{2}(P_A^0 + P_A^3) $.

For all the spinless point-particle backgrounds discussed earlier this identification can be pushed to higher orders, including the quantum long-range effects for the wave scattering as described in the introduction. For their classically spinning counterpart, the correspondence is an open question beyond leading order except for low-spin wave scattering.\footnote{While for the wave scattering of minimally coupled massless scalar fields there is evidence that the Compton four-point amplitude captures correctly the classical physics of Kerr, this is not the case for the scattering of higher-spin massless fields on top of such background \cite{Bautista:2021wfy,Bautista:2022xxx}. Therefore such classically spinning backgrounds can be considered as point-like only for some applications, for example like the scalar wave scattering considered in this paper.  It would be nice to understand how to generalize the amplitude approach to correctly describe the physics of Kerr black holes, perhaps by extending the worldsheet proposal of \cite{Guevara:2020xjx}. A similar conclusion is valid for other spinning backgrounds such as the gyraton and other charged spinning backgrounds, but this hasn't been verified explicitly to date. }

\section{Celestial wave scattering on backgrounds}\label{sec:celestialwavescattering}

Can we give a holographic interpretation for scattering on non-trivial backgrounds with flat asymptotics? 
In celestial holography the basic observables are celestial amplitudes whose asymptotic states are in a boost basis rather than a momentum basis. They contain the same information as the momentum-space S-matrix but make SL(2,$\mathbb{C}$) Lorentz symmetry instead of translation symmetry manifest. 
General curved backgrounds break these global symmetries of the S-matrix. For the particle-like backgrounds discussed in section~\ref{sec:celestialwavescattering} all two-point amplitudes can, however, be related to Lorentz invariant four-point amplitudes which can, in turn, be expressed as (flat space) celestial amplitudes. These compute overlaps between asymptotic past and future states with definite boost weight rather than energy which we now briefly review.

Focusing on massless particles, we parametrize their on-shell momenta $p_i^2=0$ as 
\begin{equation}\label{eq:pmu}
    p^\mu_i=\eta_i \omega_i q^\mu_i(z_i,\bar z_i)
\end{equation}
where $\eta_i=-1$ ($\eta_i=+1$) for incoming (outgoing) particles, $\omega_i$ are the particle frequencies and the null vector $q_i$ is directed towards a point $(z_i,\bar z_i)$ on the celestial sphere\footnote{Strictly speaking this parametrization flattens the sphere to a plane.}
\begin{equation}\label{eq:qmu}
 q_i^\mu=(1+|z_i|^2,z_i+\bar z_i, i(\bar z_i-z_i),1-|z_i|^2)
\end{equation}
where $|z_i|^2\equiv z_i \bar z_i$ and we define $z_{ij}\equiv z_i-z_j$.
Given a momentum-space amplitude $\mathcal A_n(p_1,\dots,p_n)$
we obtain the corresponding celestial amplitude by a Mellin transform with respect to the external particle frequencies~\cite{Pasterski:2016qvg}
\begin{equation}\label{eq:Atilde}
 \widetilde{\mathcal A}_n(\Delta_1,q_1,\dots ,\Delta_n,q_n):=\prod_{i=1}^n\left(\int_0^\infty d\omega_i \omega_i^{\Delta_i-1}\right) \mathcal A_n(p_1,\dots,p_n)\,,
\end{equation} 
and we will henceforth omit the dependence on the $q_i$ in the argument of the celestial amplitude to avoid cumbersome notation.
This transformation prepares scattering amplitudes with conformal primary wavefunctions 
\begin{equation}
\badat{2}
\phi^\eta_\Delta(x; q)  \equiv \int_0^\infty d\omega \omega^{\Delta-1} e^{i \eta \omega q\cdot x-\varepsilon \omega} =\frac{(-i\eta )^\Delta\Gamma(\Delta)}{(-q\cdot x-i\eta \varepsilon)^\Delta}
 \eadat
\end{equation}
rather than plane waves as the asymptotic wavefunctions. 
This recasts four-dimensional scattering amplitudes as two-dimensional correlation functions on the celestial sphere\footnote{We omit here the SL(2,$\mathbb C$) spin $J$ since we are only considering $J=0$.}
\begin{equation}
    \widetilde{\mathcal{A}}_{n}(\Delta_1,\dots,\Delta_n)=\langle \mathcal O^{\eta_1}_{\Delta_1}(z_1,\bar z_1)\dots \mathcal O^{\eta_n}_{\Delta_n}(z_n,\bar z_n)\rangle_{\rm CCFT}\,.
\end{equation}

For special values of the conformal dimensions $\Delta \in 1-\mathbb Z_{>0}$ celestial operators become conformally soft~\cite{Donnay:2018neh}, and when inserted into celestial amplitudes they give rise to the celestial analogue of soft factorization theorems~\cite{Fan:2019emx,Nandan:2019jas,Pate:2019mfs,Adamo:2019ipt,Puhm:2019zbl,Guevara:2019ypd,Fotopoulos:2020bqj}. These can in turn can be interpreted as CCFT Ward identities for asymptotic symmetries in gauge theory and gravity in asymptotically flat space~\cite{Cheung:2016iub,Donnay:2018neh,Donnay:2020guq,Pano:2021ewd,Freidel:2021ytz}. The semi-infinite tower of conformally soft operators is organized in CCFT multiplets~\cite{Pasterski:2021fjn} and obeys interesting holographic symmetry algebras~\cite{Guevara:2021abz,Strominger:2021lvk,Himwich:2021dau}.

The goal of this section is to directly express the two-particle scattering on the background as CCFT observables. This amounts to Mellin transforming the momentum-space amplitudes $\mathcal A_2^{(1)}(p_1,p_2)$ and $\mathcal M_2^{(1)}(p_1,p_2)$ considered in section \ref{sec:scalarwavescattering} in the external frequencies $\omega_1,\omega_2$ to the celestial amplitudes
$\widetilde{\mathcal A}_2^{(1)}(\Delta_1,\Delta_2)$ and $\widetilde{\mathcal M}_2^{(1)}(\Delta_1,\Delta_2)$.
In the following sections \ref{sec:Celestial_static}~-~\ref{sec:Celestial_spinning} we compute celestial amplitudes for two massless scalars scattering off the Kerr-Schild backgrounds discussed in section~\ref{sec:scalarwavescattering}.
We consider in section \ref{sec:Conformallysoft} their conformally soft limits. In section \ref{sec:CelestialCorrelator} we show that for shockwave backgrounds the celestial two-point amplitudes can be interpreted as standard CFT three-point functions of two massless asymptotic states and a conformal primary shockwave operator.

\subsection{Coulomb and Schwarzschild}\label{sec:Celestial_static}
We consider massless scalars with momenta parametrized as in~\eqref{eq:pmu} and take particle one to be incoming ($\eta_1=-1$) and particle two to be outgoing ($\eta_2=+1$).
Their celestial two-point amplitude in the Coulomb background is 
\begin{equation}\label{eq:celestial2ptCoulomb}
       \widetilde{\mathcal A}_{2,Coulomb}^{(1)}(\Delta_1,\Delta_2)=\pi eQ \frac{1}{|z_{12}|^2} \left(\frac{1+|z_1|^2}{1+|z_2|^2}\right)^{\Delta_2-1} \mathcal I(\Delta_1+\Delta_2-2),
\end{equation}
while in the Schwarzschild geometry it is given by
\begin{equation}\label{eq:celestial2ptSchw}
    \widetilde{\mathcal M}_{2,Schwarzschild}^{(1)}(\Delta_1,\Delta_2)=8\pi^2 G M \frac{1}{|z_{12}|^2} (1+|z_2|^2)\left(\frac{1+|z_1|^2}{1+|z_2|^2}\right)^{\Delta_2} \mathcal I(\Delta_1+\Delta_2-1)\,.
\end{equation}
Here we have defined the integral
\begin{equation}\label{eq:Idef}
\badat{2}
 \mathcal I(s)&\equiv \int_0^\infty d\omega \omega^{s-1}\,,
 \eadat
\end{equation}
which we may regard as a generalized distribution $\mathcal I(s)\equiv 2\pi \boldsymbol{\delta}(is)$ defined in~\cite{Donnay:2020guq} that reduces to the ordinary Dirac delta function $ I(s)=2\pi \delta({\rm Im}(s))$ when ${\rm Re}(s)=0$.

When the conformal dimensions of the massless scalars are on the principal continuous series of the Lorentz group, that is $\Delta_i\in 1+i\lambda_i$ with $\lambda_i\in \mathbb{R}$, the integral  $\mathcal I(\Delta_1+\Delta_2-2)=\delta(\lambda_1+\lambda_2)$ can be interpreted as a distribution. Thus scattering massless scalars in a Coulomb potential leads to a well-defined celestial amplitude~\eqref{eq:celestial2ptCoulomb}. 
In general, Mellin transforms of momentum-space amplitudes are not convergent even in flat space,
and so it comes as no surprise that neither is the integral in~\eqref{eq:celestial2ptSchw}. Indeed, the extra factor of $p_2^0=\omega_2(1+|z_2|^2)$ in the gravitational amplitude results in a shift in conformal dimension $\Delta_2\to\Delta_2+1$ in \eqref{eq:celestial2ptSchw} compared to the Coulomb background \eqref{eq:celestial2ptCoulomb}. This spoils the marginal convergence of \eqref{eq:Idef} when $\Delta_i\in 1+i\mathbb R$ for the Schwarzschild geometry.

Note that while we have picked a specific reference direction $u^\mu=(1,0,0,0)$ for the backgrounds we can easily generalize the celestial two-point amplitudes to any constant timelike $u^\mu$. This is achieved by replacing $1+|z_i|^2\to  q_i\cdot u$ in \eqref{eq:celestial2ptCoulomb} and  \eqref{eq:celestial2ptSchw}. 
What is striking is the dependence on the relative separation of the celestial coordinates $z_{ij}$. Unlike celestial (low-point) amplitudes in flat space which have kinematic delta functions in the $z_{ij}$, celestial two-point amplitudes in the Coulomb and Schwarzschild backgrounds have power-law behavior! We will return to this point in section~\ref{sec:CelestialCorrelator}.

\subsection{Shockwaves}\label{sec:Celestial_shock}
The ultra-boosted Coulomb and Schwarzschild backgrounds have the factors of $p_i^0$ in the scalar two-point amplitudes replaced by $p_i^-=\frac{1}{2}(p^0-p^3)$. In the parametrization~\eqref{eq:pmu} this amounts to $1+|z_i|^2\to |z_i|^2$. The celestial two-point amplitude for scattering massless scalars in the electromagnetic shockwave background is
\begin{equation}\label{eq:celestial2ptEMshock}
   \widetilde{\mathcal A}_{2,shockwave}^{(1)}(\Delta_1,\Delta_2)=\pi e Q \frac{1}{|z_{12}|^2} \left(\frac{|z_1|^2}{|z_2|^2}\right)^{\Delta_2-1} \mathcal I(\Delta_1+\Delta_2-2),
\end{equation}
while in the Aichelburg-Sexl geometry we find 
\begin{equation}\label{eq:celestial2ptgravityshock}
    \widetilde{\mathcal M}_{2,shockwave}^{(1)}(\Delta_1,\Delta_2)=16\pi^2 GP^+ \frac{1}{|z_{12}|^2} |z_2|^2\left(\frac{|z_1|^2}{|z_2|^2}\right)^{\Delta_2} \mathcal I(\Delta_1+\Delta_2-1),
\end{equation}
with the integrals $\mathcal I(s)$ defined in~\eqref{eq:Idef}. Thus we also get a well-defined celestial two-point amplitude in the background of an electromagnetic shockwave, which has distributional support when the conformal dimensions lie on the principal continuous series. The two-point amplitude in the gravitational shockwave background has again shifted conformal dimensions. By replacing $|z_i|^2\to  \tfrac{1}{2}q_i\cdot k$  we can generalize these celestial correlators from $k^\mu=(1,0,0,1)$ to any constant null direction of the shockwave backgrounds.

\subsection{Spinning shockwaves}\label{sec:Celestial_spinningshock}
Celestial wave scattering on backgrounds with classical spin yields several new features. Defining the integral
\begin{equation}\label{eq:Iprimedef}
\badat{2}
    \mathcal I'(s)&\equiv i\pi\alpha^{1+s}\int_0^\infty d\omega \omega^s H_{-1}^{(2)}(2 \alpha \omega)\,,
\eadat
\end{equation}
we can express the amplitude for massless scalars scattering on the spinning electromagnetic shockwave as
\begin{equation}
\label{eq:celestial2ptspinningEMshock}
   \widetilde{\mathcal A}_{2,spinwave}^{(1)}(\Delta_1,\Delta_2)=\pi e Q \frac{ a^{2-\Delta_1-\Delta_2}}{|z_{12}|^{\Delta_1+\Delta_2}} \left(\frac{|z_1|^2}{|z_2|^2}\right)^{\frac{\Delta_2-\Delta_1}{2}}\mathcal I'(\Delta_1+\Delta_2-2)\,,
\end{equation}
while for the spinning shockwave metric we have
\begin{equation}\label{eq:celestial2ptspinninggraityshock}
\widetilde{\mathcal M}_{2,spinwave}^{(1)}(\Delta_1,\Delta_2)=16\pi^2 GP^+ \frac{a^{1-\Delta_1-\Delta_2}}{|z_{12}|^{\Delta_1+\Delta_2+1}}|z_2|^2 \left(\frac{|z_1|^2}{|z_2|^2}\right)^{\frac{\Delta_2-\Delta_1+1}{2}}\mathcal I'(\Delta_1+\Delta_2-1)\,.
\end{equation}
For the integral in~\eqref{eq:Iprimedef} we have $\alpha=a |z_{12}|\frac{|z_1|}{|z_2|}$ and thus, with ${\rm Re}(\alpha)\geq 0$ and ${\rm Im}(\alpha)=0$, we obtain 
\begin{equation}
    \mathcal{I}'(s)=-\frac{i\pi}{2}\frac{\Gamma(1+s/2)}{\Gamma(1-s/2)}\left(1+i\cot(\pi s/2)\right)\,, \quad 0<{\rm Re}(s)<\tfrac{1}{2}\,.
\end{equation}

This result is quite remarkable. Celestial two-point amplitudes on spinning electromagnetic shockwave backgrounds have support away from the principal series unlike their non-spinning counterparts.\footnote{Strictly speaking the principal series $\Delta_i=1+i\mathbb{R}$ is not included in the range of the integral in~\eqref{eq:Iprimedef}. Nevertheless, we can analytically continue the result. In the $s\to 0$ limit we recover the behavior of the amplitude in the non-spinning shockwave background.} Moreover, while for the non-spinning gravitational backgrounds the Mellin integral diverges, we find well-defined celestial amplitudes in the spinning shockwave geometry!
We attribute this to the fact that the classical spinning solution provides a finite size $a$, and therefore the UV behaviour of the scattering amplitude is better than what we would expect from the scattering of higher-spin particles \cite{Hinterbichler:2017qcl}.

\subsection{$\sqrt{\text{Kerr}}$ and Kerr}\label{sec:Celestial_spinning}
The celestial two-point amplitude in the background of a spinning charge is 
\begin{equation}
\badat{2}
   \widetilde{\mathcal A}_{2,\sqrt{Kerr}}^{(1)}(\Delta_1,\Delta_2)&=\pi e Q \frac{a^{2-\Delta_1-\Delta_2}}{|z_{12}|^2}\left(\frac{1+|z_1|^2}{1+|z_2|^2}\right)^{\Delta_2-1} \frac{(1+|z_2|^2)^{\Delta_1+\Delta_2-2}}{[2(|z_1|^2-|z_2|^2)]^{\Delta_1+\Delta_2-2}}\\
   &\times \left[\mathcal I''_{even}(\Delta_1+\Delta_2-2)-\frac{z_1\bar z_2-z_2\bar z_1}{|z_1|^2-|z_2|^2} \mathcal I''_{odd}(\Delta_1+\Delta_2-2)\right]\,,
   \eadat
\end{equation}
while in the Kerr geometry it is given by
\begin{equation}
\badat{2}
   \widetilde{\mathcal M}_{2,Kerr}^{(1)}(\Delta_1,\Delta_2)&=8\pi^2 GM \frac{a^{1-\Delta_1-\Delta_2}}{|z_{12}|^2}\left(\frac{1+|z_1|^2}{1+|z_2|^2}\right)^{\Delta_2}
  \frac{(1+|z_2|^2)^{\Delta_1+\Delta_2}}{[2(|z_1|^2-|z_2|^2)]^{\Delta_1+\Delta_2-1}}\\
   &\times   \left[\mathcal I''_{even}(\Delta_1+\Delta_2-1)-\frac{z_1\bar z_2-z_2\bar z_1}{|z_1|^2-|z_2|^2} \mathcal I''_{odd}(\Delta_1+\Delta_2-1)\right]\,.
 \eadat
\end{equation}
Here we are again considering $a^\mu=(0,0,0,a)$ and we defined the integrals  
\begin{equation}
    \mathcal I''_{even}(s)\equiv \alpha^s \int_0^\infty d\omega \omega^{s-1} \cosh(\alpha \omega)\,, \quad    \mathcal I''_{odd}(s)\equiv \alpha^s \int_0^\infty d\omega \omega^{s-1} \sinh(\alpha \omega)\,,
\end{equation}
where $\alpha = 2a \frac{|z_1|^2-|z_2|^2}{1+|z_2|^2}$. These integrals are well-defined only for imaginary spin parameter~$a$, or ${\rm Re(\alpha)}=0$, for which we find
\begin{equation}
    \mathcal I''_{even}(s)=i^s\cos\left(\frac{\pi s}{2}\right)\Gamma(s)\,, \quad \mathcal I''_{odd}(s)=i^s\sin\left(\frac{\pi s}{2}\right)\Gamma(s)\,, \quad 0<{\rm Re}(s)<1\,.
\end{equation}
We then analytically continue the final result for the celestial amplitudes to real value of $a$. 
Remarkably, the Mellin transform gives a well-defined celestial amplitude with support away from the principal series both in gauge theory and gravity.

It is worth commenting here on the relation with the spinning shock-wave results. While the original $\sqrt{\text{Kerr}}$ and Kerr metrics are related to the electromagnetic and gravitational spinning shockwaves by an infinite boost along the direction of $a^{\mu}$ \cite{Ferrari:1990tzs}, it is unclear how such a boost limit is encoded in the two-point amplitude discussed above. As we will see later in section \ref{sec:CelestialCorrelator}, it is expected that we need a combination of conformal primaries to represent massive background solutions while a single operator is sufficient for shockwave metrics.

\subsection{Conformally soft limits}\label{sec:Conformallysoft}

Let us discuss the behavior of celestial two-point amplitudes on particle-like backgrounds in the conformally soft limit. In the momentum basis this corresponds to probing the backgounds by low-energy scatterers. Taking the soft limit $\omega_1\to0$ in the momentum-space amplitudes implies also $\omega_2\to0$ by momentum-conservation. Energetically soft limits of momentum-space amplitudes map to poles in celestial amplitudes. To see this note that the Mellin integral up to some cut-off $\omega \ll \omega_*$ takes the form
\begin{equation}
    \int_0^{\omega_*}d\omega \omega^{s-1}=\frac{\omega_*^{s}}{s}, \quad {\rm Re}(s)>0\,.
\end{equation}
Let's first look at the static amplitudes 
\begin{equation}
\badat{2}
    \widetilde{\mathcal A}_{2}^{(1)}(\Delta_1,\Delta_2)&=\pi e Q \frac{1}{|z_{12}|^2} \left(\frac{q_1\cdot u}{q_2\cdot u}\right)^{\Delta_2-1} \mathcal I(\Delta_1+\Delta_2-2)\,,\\
    \widetilde{\mathcal M}_{2}^{(1)}(\Delta_1,\Delta_2)&=8\pi^2  G r_0 \frac{1}{|z_{12}|^2} (q_2\cdot u) \left(\frac{q_1\cdot u}{q_2\cdot u}\right)^{\Delta_2} \mathcal I(\Delta_1+\Delta_2-1)\,,
\eadat
\end{equation}
corresponding to Coulomb and Schwarzschild (with $r_0=M$) for constant time-like $u^\mu$ and to their ultraboost limits for light-like $u^\mu$ (with $r_0=P^+$, the lightcone energy, for the Aichelburg-Sexl shockwave). 
The energetically soft limit maps to a pole at $\Delta_1+\Delta_2=2$ such that
\begin{equation}\label{eq:A2confsoft}
   \lim_{\Delta_1+\Delta_2\to 2} (\Delta_1+\Delta_2-2) \widetilde{\mathcal A}_{2}^{(1)}(\Delta_1,\Delta_2)=  \lim_{\Delta_1+\Delta_2\to 2}\pi e Q \frac{1}{|z_{12}|^2}\left(\frac{q_1\cdot u}{q_2\cdot u}\right)^{\frac{\Delta_2-\Delta_1}{2}}.
\end{equation}
For the Schwarzschild and Aichelburg-Sexl geometry we have $s=\Delta_1+\Delta_2-1$ so that the energetically soft limit maps to a pole at $\Delta_1+\Delta_2=1$ and we get
\begin{equation}
   \lim_{\Delta_1+\Delta_2\to 1} (\Delta_1+\Delta_2-1) \widetilde{\mathcal M}_{2}^{(1)}(\Delta_1,\Delta_2)=\lim_{\Delta_1+\Delta_2\to 1}8\pi^2 Gr_0 \frac{(q_2\cdot u)}{|z_{12}|^2}\left(\frac{q_1\cdot u}{q_2\cdot u}\right)^{\frac{\Delta_2-\Delta_1+1}{2}}\,.
\end{equation}

The same behavior is obtained for the Kerr and spinning shockwave backgrounds. 
In particular from \eqref{eq:celestial2ptspinningEMshock} and \eqref{eq:celestial2ptspinninggraityshock} we see that in these ``simultaneous'' conformally soft limits the dependence on the spin drops out! This can be traced back to the spin and the energy appearing in the combination $(a \omega)^s$.

\subsection{Celestial shockwave correlators}\label{sec:CelestialCorrelator}

Finally, let us return to the question raised at the beginning of this section. Can we extend the map between bulk scattering processes and celestial CFT correlators beyond flat spacetimes? Non-trivial backgrounds such as the ones considered in this work explicitly break Lorentz symmetry. Nevertheless, the electromagnetic and gravitational shockwave backgrounds do so in a very mild manner, and in fact can themselves be interpreted as conformal primary backgrounds. Interestingly, we can use them to construct celestial correlators that transform covariantly under SL(2,$\mathbb C$) and are free of the kinematic delta functions that celestial low-point amplitudes in flat space are plagued by. In the case of gravity, this requires an analytic continuation off the principal series $\Delta\in1+i\mathbb R$ which renders the divergent Mellin integrals finite.

From the perspective of the celestial CFT, the electromagnetic shockwave corresponds to a generalised conformal primary vector of $\Delta=0, J=0$, while the gravitational shockwave is a generalised conformal primary metric of $\Delta=-1, J=0$ \cite{Pasterski:2020pdk}. Consequently, we can view the propagator of massless scalars in a shockwave background as a three-point function of two conformal asymptotic states and a shockwave primary in the celestial CFT. 

To make this explicit consider the three-point function for two massless scalars and an off-shell photon of momentum $p$,
\begin{equation}
    \mathcal{A}_{3;\mu}(p_1,p_2, p) = e (2\pi)^4 \delta^{(4)}(p_1+p_2 +p) (p_{1\mu}-p_{2\mu})~.
\end{equation}
This quantity is essentially the momentum space form factor, $\bra{p_1}\bar{j}_\mu(p)\ket{p_2}$ of the electromagnetic current ${j}_\mu=ie (\phi\partial_\mu {\phi}^\ast-{\phi}^\ast\partial_\mu \phi)$. 
To go from the on-shell plane-wave momentum eigenstates for the scalars to conformal primary wavefunctions we simply carry out the Mellin transform, while to deal with the off-shell plane-wave basis we make use of the fact that the electromagnetic shockwave wavefunction can be written as~\cite{Pasterski:2020pdk}
\begin{equation}
\badat{2}
    A^{\text{sw}}_{0,0;\mu}(x;q)&=- q_{\mu}  \log (x^2)  \delta(q\cdot x)\\
&=8\pi^2q_{\mu}\int \frac{d^4 p}{(2\pi)^4} \frac{\delta(p\cdot  q)}{p^2} e^{ip\cdot x}\,.
\eadat
\end{equation}
Thus, defining a shockwave null vector $q_{\text{sw}}$ with components
\begin{equation}
\label{eq:sw_mom}
q_{\text{sw}}^\mu=(1+|z_{\text{sw}}|^2, z_{\text{sw}}+\bar{z}_{\text{sw}},i(\bar{z}_{\text{sw}}-z_{\text{sw}}) , 1-|z_{\text{sw}}|^2)~.
\end{equation}
 we find for the transformed  amplitude
\begin{equation}
\badat{2}
\widetilde{\mathcal{A}}_3(\Delta_1, \Delta_2, \Delta_{\text{sw}}=0)&\equiv\frac{2}{(2\pi)^2}\int d\omega_1 d\omega_2 \omega_1^{\Delta_1-1}\omega_2^{\Delta_2-1}\int d^4 p \frac{\delta(p\cdot  q_{\text{sw}})}{p^2}  q_{\text{sw}}\cdot \mathcal{A}_{3}(p_1,p_2, p)\\
&=\frac{e(2\pi)^3\boldsymbol{\delta}(i(\Delta_1+\Delta_2-2))}{|z_{12}|^{\Delta_1+\Delta_2}|z_{1\text{sw}}|^{\Delta_1-\Delta_2}|z_{2\text{sw}}|^{\Delta_2-\Delta_1}}
\eadat
\end{equation}
where $z_{i\text{sw}}\equiv z_i-z_{\text{sw}}$ and $\boldsymbol{\delta}$ reduces to the ordinary Dirac delta function for $\Delta_i \in 1+i\mathbb R$. We see that the coordinate dependence is simply that of standard two-dimensional CFTs consistent with $\Delta_{\text{sw}}=0$! This amplitude also agrees, up to a factor of the background charge, with the two-point function in the shockwave background which can be seen upon substituting \eqref{eq:sw_mom} into the generalized version of \eqref{eq:celestial2ptEMshock}. Thus we see that, in celestial coordinates, two-point functions in the shockwave background can be interpreted as three-point correlation functions between two asymptotic states and a shockwave operator. 

Repeating the argument for the gravitational case we write the shockwave metric as\footnote{Since we are now computing the quantum amplitude with perturbative expansion $g_{\mu\nu}=\eta_{\mu\nu}+\kappa h_{\mu\nu}$ we have to divide by $\kappa$ in \eqref{eq:hsw3pt}.}~\cite{Pasterski:2020pdk}
\begin{equation}\label{eq:hsw3pt}
\badat{2}
    h^{\text{sw}}_{-1,0;\mu\nu}(x,q)&=-\frac{1}{\kappa}q_{\mu} q_{\nu}  \log (x^2)\delta(q \cdot x)\\
&=8\pi^2 \frac{1}{\kappa}q_{\mu} q_{\nu}\int \frac{d^4 p}{(2\pi)^4} \frac{\delta(p\cdot  q)}{p^2} e^{ip\cdot x}~
\eadat
\end{equation}
and so transform the three-point amplitude for two massless scalars and an off-shell graviton\footnote{This quantity is the momentum space form factor of the scalar stress-energy tensor $T_{\mu\nu}=\partial_{\mu} \phi \partial_\nu \phi-\tfrac{1}{2}g_{\mu\nu} \partial^\lambda \phi \partial_\lambda \phi$. Such a quantity is not a good observable in gravity, however once we contract with the conformal primary wavefunction it becomes the matrix element of the integrated quantity $\int  d^4x \sqrt{-g} f(x) q^\mu q^\nu T_{\mu\nu}(x)$.} 
\begin{equation}
     \mathcal{M}_{3;\mu\nu}(p_1,p_2,p)=-\kappa(2\pi)^4 \delta^{(4)}(p_1+p_2-p) \big[p_{1\mu}p_{2\nu}-\tfrac{1}{2}\eta_{\mu\nu}(p_1\cdot p_2)\big]
\end{equation}
into the celestial three-point function
\begin{equation}
\badat{2}
\widetilde{\mathcal{M}}_3(\Delta_1,  \Delta_2, \Delta_{\text{sw}}=-1)&\equiv\frac{2}{(2\pi)^2}\int d\omega_1 d\omega_2 \omega_1^{\Delta_1-1}\omega_2^{\Delta_2-1}\int  d^4 p \frac{\delta(p\cdot  q_{\text{sw}})}{p^2}  q_{\text{sw}}^\mu q_{\text{sw}}^\nu \mathcal{M}_{3;\mu\nu}(p_1,p_2, p)\\
&= \frac{(2\pi)^2\mathcal{I}(\Delta_1+\Delta_2-2)}{|z_{12}|^{2}|z_{1\text{sw}}|^{-2\Delta_2}|z_{2\text{sw}}|^{2\Delta_2-2}}~.
\eadat
\end{equation}
This quantity does not have the form of a CFT correlator and suffers from the usual divergence for operators on the principle series. However, if we continue away from these values such that Re$(\Delta_1+\Delta_2)=1$, we can write this as 
\begin{equation}
\widetilde{\mathcal{M}}_3(\Delta_1 ,\Delta_2, \Delta_{\text{sw}}=-1)= \frac{(2\pi)^3\boldsymbol{\delta}(i(\Delta_1+\Delta_2-1))}{|z_{12}|^{\Delta_1+\Delta_2+1}|z_{1\text{sw}}|^{\Delta_1-\Delta_2-1}|z_{2\text{sw}}|^{\Delta_2-\Delta_1-1}}
\end{equation}
which has the correct coordinate dependence. This agrees, up to a factor of the background charge which in this case is the light-cone energy, with the two-point function \eqref{eq:celestial2ptgravityshock} in the Aichelburg-Sexl geometry.

The case of spinning shockwaves is particularly interesting as the transformation to the celestial basis is well-defined even in the gravitational case. The conformal vector primary and conformal metric primary are related to the off-shell plane wave basis by the transform 
\begin{equation}
    A^{\text{ssw}}_{0,0;\mu}(X,q)=q_{\mu} \phi^{\text{ssw}}(X,q),~~~~ h^{\text{ssw}}_{-1,0;\mu}(X,q)=q_{\mu} q_{\nu} \phi^{\text{ssw}}(X,q),
\end{equation}
with 
\begin{equation}
    \phi^{\text{ssw}}(X,q)=-\delta(q\cdot X) \log(X^2-a^2)=4\pi^3 i a\int d^4p \frac{\delta(p\cdot q)}{|p|} H^{(2)}_{-1}(a|p|) e^{i p\cdot X}.
\end{equation}
Following similar steps as before, and defining a null vector $q_{\text{ssw}}$ parameterised by complex parameters $z_{\text{ssw}}$, we find the celestial three-point functions
\begin{equation}\label{eq:Celestial3ptEMspinning}
    \widetilde{\mathcal{A}}_3(\Delta_1, \Delta_2, \Delta_{\text{ssw}}=0)=\frac{e(2\pi)^2 a^{2-\Delta_1-\Delta_2}\mathcal I'(\Delta_1+\Delta_2-2)}{|z_{12}|^{\Delta_1+\Delta_2}|z_{1\text{ssw}}|^{\Delta_1-\Delta_2}|z_{2\text{ssw}}|^{\Delta_2-\Delta_1}}
\end{equation}
for the electromagnetic spinning shockwave and
\begin{equation}\label{eq:Celestial3ptgravityspinning}
\widetilde{\mathcal{M}}_3(\Delta_1, \Delta_2, \Delta_{\text{ssw}}=-1)=\frac{r_0(2\pi)^2  a^{1-\Delta_1-\Delta_2} \mathcal I'(\Delta_1+\Delta_2-1)}{|z_{12}|^{\Delta_1+\Delta_2+1}|z_{1\text{ssw}}|^{\Delta_1-\Delta_2-1}|z_{2\text{ssw}}|^{\Delta_2-\Delta_1-1}}
\end{equation}
for the gravitational spinning shockwave, where we defined $z_{i\text{ssw}}\equiv z_i-z_{\text{ssw}}$. Again their dependence on the celestial coordinates is that of CFT three-point correlators. Comparing \eqref{eq:Celestial3ptEMspinning} and \eqref{eq:Celestial3ptgravityspinning} to the two-point functions \eqref{eq:celestial2ptspinningEMshock} and \eqref{eq:celestial2ptspinninggraityshock} in the spinning shockwave backgrounds we again find agreement.

\section{Boundary on-shell action localization}\label{sec:bdyaction}
In the previous section we made use of the fact that the classical solution to the equations of motion on AF backgrounds 
is the generating functional for tree-level correlation functions in CCFT. In AdS/CFT this role is played by the boundary on-shell action. In asymptotically flat space we, similarly, expect the on-shell action to localize near the asymptotic boundary at infinite radial distance and that it will generate correlation functions in the boundary theory.\footnote{The fact that the classical action contains all the relevant information for the tree-level scattering on a background is a well-known statement (see the nice discussion in appendix B of \cite{Adamo:2017nia} for plane-wave backgrounds). In the eikonal case, this localization has also been shown recently for point-particle backgrounds in \cite{Adamo:2021rfq}. Here we extend such analysis to derive a more general statement which include both classical ``eikonal'' and quantum contributions to the wave scattering.} For massless particles we anticipate localization on the celestial sphere at null infinity as we will now show, focusing for simplicity on the two-point function. The procedure can, however, be extended to higher functions by turning on the appropriate number of sources.

The action for a complex massless scalar $\phi$ minimally coupled to the background metric $g_{\mu \nu}$
is given by
\begin{align}
\mathcal{S} = \int \mathrm{~d}^{4} x \sqrt{-g} \phi^*(x) \DAlambert \phi(x)-\int \mathrm{~d}^{4} x \sqrt{-g} \nabla_{\mu} [\phi^*(x) g^{\mu \nu} \nabla_{\nu} \phi(x)] \,,
\end{align}
where we have already isolated the boundary term
\begin{align}
\mathcal{S}_{\text{bdy}}&\equiv-\int \mathrm{~d}^{4} x \sqrt{-g} \nabla_{\mu} [\phi^*(x) g^{\mu \nu} \nabla_{\nu} \phi(x)]\,,
\end{align}
and we must include additional source terms, $\mathcal{S}^{\text{src}}=\int d^4 x (J \phi^\ast+J^\ast \phi)$. 
We assume that all our backgrounds have an asymptotically flat spacetime region described at leading order by the Minkowski metric
\begin{align}
d s^2 = - d t^2 + d r^2 + r^2 (d\theta^2+\sin^2\theta d\varphi^2) \,.
\end{align}
We reach past and future time-like infinity by respectively taking $t\to -\infty$ and $t\to +\infty$ while keeping $r$ fixed, spatial infinity corresponds to taking $r\to \infty$ while keeping $t$ fixed, while past and future null infinity are reached by taking $r\to \infty$ while respectively holding advanced time $v=t+r$ and retarded time $u=t-r$ fixed.
The surface term in the action is an integral over the null
boundary $\mathscr{I}^- \cup \mathscr{I}^+$. Following~\cite{Fabbrichesi:1993kz}, we may define a general null hypersurface in terms of the zero of a piecewise smooth function $\tau(x)$ whose sign is chosen so that the direction of growing $\tau$ points outward.
Gauss's theorem can then be written as
\begin{equation}\label{SbdyGauss}
    \mathcal{S}_{\text{bdy}} =-\int d^4 x \sqrt{-g} \delta(\tau(x)) \left[\phi^*(x)g^{\mu\nu}\nabla_\nu \phi(x)\right] \nabla_\mu (\tau(x))\,.
\end{equation}
The contribution to~\eqref{SbdyGauss} from $\mathscr{I}^\pm$ is then obtained from
 \begin{equation}
    \tau^+(x)=-\frac{1}{v(x)}+\gamma\,, \quad  \tau^-(x)=+\frac{1}{u(x)}+\gamma\,,
\end{equation}
in the limit $\gamma\to 0$ which pushes the hypersurface to null infinity. In the asymptotic region near $\mathscr{I}^+$ we have $v(x)=u+2r$, while near $\mathscr{I}^-$ we have $u(x)=v-2r$. Evaluating the integral over $r$ by means of the $\delta$-function  we find for the boundary action
\begin{equation}\label{eq:fluxCCFT}
    \mathcal{S}_{\mathscr{I}^-\cup \mathscr{I}^+}=-\lim_{r\to \infty} \int_0^{2\pi}d\varphi \int_0^\pi d\theta \sin \theta r^2 \left[ \int_{-\infty}^{+\infty} du \left(\phi^* n^+_\mu \partial^\mu \phi\right)+\int_{-\infty}^{+\infty} dv \left(\phi^* n^-_\mu \partial^\mu \phi\right)\right]
\end{equation}
where $n^+_\mu=\partial_\mu(+\frac{1}{2}v)$ and $n^-_\mu=\partial_\mu(-\frac{1}{2} u)$.

The flux factor defined by \eqref{eq:fluxCCFT}  is what determines the two-point function for the wave scattering, as pointed out in the classical analysis of \cite{Adamo:2021rfq}. In the spirit of the scattering problem we can decompose the field $\phi$ as a superposition of the incoming and the outgoing contributions
\begin{align}
\phi(x) = \phi_{\text{in}}(x) + \phi_{\text{out}}(x) \,, \qquad  \phi_{\text{in}}(x) = e^{i p \cdot x} \,,
\end{align}
where the momentum pointing towards the celestial sphere is parametrized by $p^\mu=\omega(1,\hat p)$,
and we can focus on the in-out contributions to the scattering from the action
\begin{align}
\mathcal{S}^{\text{in/out}}_{\mathscr{I}^-\cup \mathscr{I}^+}\equiv-\sum_{\eta=\pm} \int_{\mathscr{I}^{\eta}} \mathrm{~d}^{3} x \sqrt{-g_{\text{ind}}}  \phi^*_{\text{in}}(x) n^{\nu}_\eta \partial_{\nu} \phi_{\text{out}}(x)  \,.
\label{eqn:in-out_boundary}
\end{align}
The wave equation for the scalar field can be written in terms of the scalar effective source as showed in section \ref{sec:scalarwavescattering},
\begin{align}
\eta^{\mu \nu}  \partial_{\mu} \partial_{\nu} \phi(x) &= J_{\text{eff}}(x) \,
\end{align}
where $J_{\text{eff}}$ includes the external source $J$ and the coupling of the field to the non-trivial background. 
This equation can be formally solved by using standard Green's functions methods,
\begin{align}
\phi_{\text{out}}(x) &= \int_{\mathcal{C}} \frac{d k^0}{2 \pi} \int \frac{d^3 k}{(2 \pi)^3} \frac{e^{i k \cdot x}}{(k^0)^2 - |\vec{k}|^2} \bar{J}_{\text{eff}}(k^0,\vec{k}=|\vec{k}|\hat{k}) \,,
\label{eq:phi_fullsol}
\end{align}
where the contour $\mathcal{C}$ specifying the Green's function is chosen according to the boundary conditions. We will focus on the cases of advanced and retarded Green's functions which correspond to placing all sources in the causal future or past of the field. 

We then take the large $r$ limit, holding either $v$ or $u$ fixed, to obtain the leading component of the scalar field around each null boundary: $\mathscr{I}^-$ from the advanced contribution or $\mathscr{I}^+$ from the retarded, (see e.g. \cite{Cristofoli:2021vyo,Maggiore:2007ulw}). 
The saddle point approximation then localizes the $\hat k$ integral near $\mathscr{ I}^\pm$
at $\hat k= \pm \hat x$ where $\hat x=\vec{x}/r$  and we obtain 
\begin{align}
\phi_{\text{out}}(t\mp r,r,\hat{x})  &= -c_{\pm}\frac{1}{4 \pi r} \int_{\mathbb{R}} \frac{d \omega_k}{2 \pi} \,e^{\mp i \omega_k (t \mp r)} \bar{J}_{\text{eff}}(\pm \omega_k, \omega_k \hat{x}) + \mathcal{O}\left(r^{-2}\right) \,
\label{eq:sol_effsource}
\end{align}
where $c_{+}=1$, $c_-=0$ for the retarded choice of contour and $c_{+}=0$, $c_-=1$  for the advanced. 
This is more general than the eikonal wave scattering solution in  \cite{Adamo:2021rfq}, because it also allows for quantum contributions to the scattering of waves on top of the (classical) background. 
We are now ready to compute the boundary term as a function of the momentum\footnote{Note that while in section~\ref{sec:scalarwavescattering} we used the in-in formalism labeling states by $\eta=-1$ (incoming) and $\eta=+$ (outgoing), here we are working in the in-out formalism.}
\begin{equation}
    \badat{2}
\mathcal{S}_{\mathscr{I}^+}^{\text{in/out}} (p)=&
-\frac{1}{8 \pi^2} \lim_{r \to \infty}   c_+ \int_{\mathscr{I}^+} d u  d \theta d\varphi \, \sin\theta\, e^{i \omega u} \\
&  \quad \quad\times \int_{\mathbb{R}} d \omega_k \,  (i \omega_k r-1) \bar{J}_{\text{eff}}( \omega_k,\omega_k \hat{x}) e^{-i  \omega_k u} e^{-i\omega r (\hat{p}\cdot \hat{x}-1)} 
   \eadat
\end{equation}
for the future null boundary, and
\begin{equation}
    \badat{2}
\mathcal{S}_{\mathscr{I}^-}^{\text{in/out}}(p)=&
-\frac{1}{8 \pi^2} \lim_{r \to \infty}   c_- \int_{\mathscr{I}^-} d v   d \theta d\varphi \, \sin\theta\, e^{i \omega v} \\
&  \quad \quad\times \int_{\mathbb{R}} d \omega_k \,  (i \omega_k r-1) \bar{J}_{\text{eff}}( -\omega_k,\omega_k \hat{x}) e^{-i  \omega_k v} e^{-i\omega r (\hat{p}\cdot \hat{x}+1)} 
   \eadat
\end{equation}
for the past null boundary.
The integrals over retarded and advanced time localize $\omega_k=\omega$,
\begin{equation}
\mathcal{S}_{\mathscr{I}^-\cup \mathscr{I}^+}^{\text{in/out}} (p)=  -\frac{1}{4 \pi} \lim_{r \to \infty}  \sum_{\eta=\pm} c_\eta \eta \int d \theta d \varphi \, \sin\theta\, \,  (i \omega r) \bar{J}_{\text{eff}}(\omega,\eta \omega \hat{x}) e^{-i \omega r( \hat{p}\cdot \hat{x}-\eta)}   \,,
\end{equation}
where $\eta=\pm$ for $\mathscr{ I}^\pm$. To solve the integrals over the angles we again make use of the saddle point approximation setting $\hat x=\pm \hat p$ for $\mathscr{I}^{\pm}$ which yields
\begin{equation}
\label{eq:action_localization}
\mathcal{S}_{\mathscr{I}^-\cup \mathscr{I}^+}^{\text{in/out}}(p) = \left( \frac{c_++c_-}{2}\right) \bar{J}_{\text{eff}}(\omega,\omega \hat{p}) \,.
\end{equation}
This is a very interesting result: 
the boundary term in the action localizes to the Fourier transform of the effective source evaluated at large distances along the incoming momentum. The normalisation of the result depends on the choice of contour and with the advanced or retarded prescriptions the boundary action is essentially half of the effective source. If one considers linear combinations of solutions, which would still have vanishing bulk action when all sources are at the boundary, one sees that for the antisymmetric combination, corresponding to using the causal propagator, the boundary action vanishes while the symmetric combination gives a boundary action equal to the effective source. 

The relation with the two-point function discussed in the section \ref{sec:scalarwavescattering} is then easily found. From~\eqref{eq:npointA} and using \eqref{eq:phi_fullsol} and \eqref{eq:action_localization} we find 
\begin{equation}
\badat{3}
\mathcal M_2(p_1,p_2)&=-\lim_{p_1^2\to 0}\lim_{p_2^2\to 0} p_1^2p_2^2 \frac{\delta\bar \phi_{\text{out}}(-p_1)}{\delta \bar J(p_2)} \\
&= \lim_{p_1^2\to 0}\lim_{p_2^2\to 0} p_2^2 \frac{\delta \bar{J}_{\text{eff}}(-p_1)}{\delta \bar J(p_2)}  \\
&= \lim_{p_1^2\to 0}\lim_{p_2^2\to 0} p_2^2 \frac{\delta \Big[\left(\frac{2}{c_++ c_-}\right)\mathcal{S}_{\mathscr{I}^-\cup \mathscr{I}^+}^{\text{in/out}}(-p_1)\Big]}{\delta \bar J(p_2)} \,,
\eadat
\label{eq:2Amp_fullsol}
\end{equation}
where we have set $p=p_1$ as the incoming momentum. It is important to note that all the quantities in \eqref{eq:2Amp_fullsol}, and their corresponding physical interpretation, depend implicitly on the choice of contour. This shows that, similarly to the AdS/CFT case, the boundary on-shell action is the generating functional of the two-point function. It is straightforward to extend this derivation to a complex massless scalar minimally coupled to a Kerr-Schild gauge background, which is flat by definition. In appendix \ref{app:spinningbdyaction}, we also extend this construction to a $U(1)$ gauge field minimally coupled to a gravitational Kerr-Schild background.

\section{Conformal Faddeev-Kulish dressings for backgrounds}\label{sec:KSdressing}

In the previous sections we have considered only the leading contribution to the two-point function. At higher orders in perturbation theory, infrared divergences will appear as a consequence of the non-trivial asymptotic dynamics in the presence of long-range massless interactions. This can be seen by iteratively solving the wave equations.

In the electromagnetic case \eqref{eq:em_we} for Kerr-Schild type potentials \eqref{eq:pot_KS}, where the non-linear $A^2$ term vanishes, we have
\begin{equation}
     \bar{\phi}(p)=-\frac{1}{p^2}\sum_{n=0}^\infty  \int \prod_{\ell=1}^n \frac{d^4k^{(\ell)}}{(2\pi)^4} \frac{\mathcal{A}^{(1)}_2(p,-k^{(1)})}{k^{(1)}{}^2}
     \frac{ \mathcal{A}^{(1)}_2(k^{(1)},-k^{(2)})}{k^{(2)}{}^2}\dots \frac{\mathcal{A}_2^{(1)}(k^{(n-1)},-k^{(n)})}{k^{(n)}{}^2}\bar{J}(k^{(n)})~.
\end{equation}
This solution can be re-summed in the eikonal approximation, where  we treat the integrated momenta $k^{(\ell)}$ as soft, by expanding propagators e.g. $1/(k^{(1)}+p_1)^2\to 1/2k^{(1)}\cdot p_1$ and dropping powers of $k^{(\ell)}$ in numerators. 
The connected part of the amplitude then becomes
\begin{equation}
    \mathcal{A}^{con, IR}_2(p_1,p_2)=\text{exp}\Big[e\int \frac{d^4 k}{(2\pi)^4} \frac{\bar{A}(-k)\cdot p_2}{k\cdot p_2} \Big] \mathcal{A}_2^{(1)}(p_1,p_2)~.
\end{equation}
Thus we see that the amplitude factorises into an exponential pre-factor capturing the IR divergences and a hard factor in a manner familiar from scattering amplitudes \cite{Yennie:1961ad}. Noting that 
\begin{align}
    \int \frac{d^4 k}{(2\pi)^4} \frac{\bar{A}(-k)\cdot p_2}{k\cdot p_2}&=i \int ds~ p_2^\mu A_\mu(p_2 s)
\end{align}
we see that the prefactor has the well known operator interpretation as a Wilson line for the hard particles. This result is actually slightly more general than for just Kerr-Schild backgrounds as in this approximation we would additionally neglect the $A^2$ contributions even if they were present. For electromagnetic potentials sourced by uniform velocity charged particles, that is Coulomb and shockwave potentials, the integral is given by 
\begin{align}
\label{eq:QED_IR_phase}
   e \int \frac{d^4 k}{(2\pi)^4} \frac{\bar{A}(-k)\cdot p_2}{k\cdot p_2}&=\frac{e Q  p_2\cdot u}{8\pi^3} 
    \int  \frac{d^{4-2\epsilon}{k}~\delta(u\cdot k)}{(k^2 +i 0^+)(k\cdot p_2+i 0^+)}=\frac{ieQ}{8\pi\epsilon}
\end{align}
where we introduced an $i 0^+$ prescription for the propagators and used dimensional regularisation, with the corresponding parameter $\epsilon$, to remove the IR divergence. 
In this background, as the IR divergences are independent of the momenta, they are invisible to the Mellin transform and so the two-point function in the celestial basis can be similarly factorised into a divergent soft factor and an IR finite hard factor
\begin{align}
    \widetilde{\mathcal{A}}^{IR}_2(\Delta_1,\Delta_2)&=\int_0^\infty \prod_{i=1}^2 d\omega_i \omega_i^{\Delta_i-1} \mathcal{A}_2^{con,IR}(\omega_1,\omega_2, q_1,q_2)=\widetilde{\mathcal{A}}_2^{soft} \widetilde{\mathcal{A}}_2^{(1)}~.
\end{align}
with $\widetilde{\mathcal{A}}_2^{soft}=e^{i Q e/8\pi \epsilon}$. As we will see below, this factorisation can be understood in terms of the conformal hard-soft factorisation of scalar QED amplitudes studied in \cite{Arkani-Hamed:2020gyp}. 

While the electromagnetic case is essentially trivial this is not so for gravitational backgrounds where there is non-trivial momentum dependence in the IR divergent factor. After iteratively solving the wave equation \eqref{eq:grav_scal_weq}, again considering Kerr-Schild backgrounds or simply neglecting non-linear terms, 
\begin{equation}
    \bar{ \phi}(p)=-\frac{1}{p^2}\sum_{n=0}^\infty  \int \prod_{\ell=1}^n \frac{d^4k^{(\ell)}}{(2\pi)^4} \frac{\mathcal{M}'_2(p,-k^{(1)})}{k^{(1)}{}^2}
     \frac{ \mathcal{M}'_2(k^{(1)},-k^{(2)})}{k^{(2)}{}^2}\dots \frac{\mathcal{M}'_2(k^{(n-1)},-k^{(n)})}{k^{(n)}{}^2}\bar{J}(k^{(n)})~
\end{equation}
where we defined an off-shell tree-level amplitude $\mathcal{M}'_2(p,k)=-\bar{h}^{\mu\nu}(p+k)\big[p_\mu k_\nu -\tfrac{1}{2}\eta_{\mu\nu}(p\cdot k +k^2)\big]$, and taking the eikonal limit, the connected part can again be written as a prefactor times the tree amplitude
\begin{equation}
    \mathcal{M}^{con, IR}_2(p_1,p_2)=\text{exp}\Big[-\int \frac{d^4 k}{(2\pi)^4} \frac{\bar{h}^{\mu\nu}(-k) p_{2\mu}p_{2\nu}}{2k\cdot p_2} \Big] \mathcal{M}_2^{(1)}(p_1,p_2)~.
\end{equation}
Again the exponential prefactor has exactly the form of a Wilson line, familiar from the eikonal approximation of scattering amplitudes in gravity \cite{Naculich:2011ry, Melville:2013qca, White:2011yy}. 
For the linearized geometry describing uniform velocity particles, which includes linearized Schwarzschild and Aichelburg-Sexl geometries, the metric perturbation in harmonic gauge \eqref{eq:lin_hg_eeq} implies an IR divergent prefactor 
\begin{align}
\label{eq:IR_grav}
   - \int \frac{d^4 k}{(2\pi)^4} \frac{\bar{h}^{\mu\nu}(-k) p_{2\mu}p_{2\nu}}{2k\cdot p_2}&=-\frac{(p_{2} \cdot u)^2 Gr_0}{\pi^2} \int  \frac{d^{4-2\epsilon}{k}~\delta(u\cdot k)}{(k^2 +i 0^+)(k\cdot p_2+i 0^+)}=-\frac{i(p_2\cdot u)G r_0}{\epsilon}~
\end{align}
where $r_0$ is a constant corresponding to the mass $M$ in the case of  Schwarzschild or lightcone energy $P^+$ in the case of the shockwave. 
 Thus we see that the all-order amplitudes in non-trivial gravitational backgrounds can also be split into a hard part and a IR divergent phase. 
 
 The same splitting occurs in the celestial basis but in this case the soft prefactor becomes an operator acting on the amplitude. In the case of the shockwave, using the null momentum parameterisation \eqref{eq:pmu} and taking $u^\mu=q^\mu$ as in \eqref{eq:sw_mom}, we can write the celestial two-point amplitude as 
\begin{align}
    \widetilde{\mathcal{M}}^{IR}_2(\Delta_1, \Delta_2)
&   = \widetilde{\mathcal{M}}_2^{soft}\widetilde{\mathcal{M}}_2^{(1)}(\Delta_1, \Delta_2)
\end{align}
with 
 \begin{equation}
 \label{eq:IR_grav_cel}
   \widetilde{\mathcal{M}}^{soft}_2  =\text{exp}\Big[-\frac{i G P^+ }{\epsilon} \left(\eta_1 e^{\partial_{\Delta_1}} |z_{1\text{sw}}|^2 -\eta_2 e^{\partial_{\Delta_2}}|z_{2\text{sw}}|^2\right)\Big]
 \end{equation}
where we see the appearance of the dimension shifting operators $e^{\partial_{\Delta_k}}$. 
For particle-like backgrounds, we can derive a dressing for the background states which removes these divergent terms at all orders in perturbation theory in the two-point function calculation by using the matching to four-point amplitudes discussed in section \ref{sec:eikonal}. 

\subsection{QED amplitudes}
The IR divergences in QED amplitudes have long been been known to exponentiate \cite{Yennie:1961ad, Weinberg:1965nx}.  Let us denote by $p_1$, $p_3=P_A$ the incoming momenta and  $p_2$, $p_4=P_B$ the outgoing momenta. We take the momenta $p_1$ and $p_2$ to be massless probe particles while $P_A$ and $P_B$ are particles generating the non-trivial background.\footnote{Here, and below, we take all momenta to be positive energy and explicitly write the factors $\eta=\pm$ previously included in the momenta.} The S-matrix element can be written as
\begin{align}
\langle p_2 P_B |S|p_1 P_A \rangle = W_{4,\text{QED}} {\langle p_2 P_B |S| p_1  P_A \rangle}_{hard} \,,
\end{align}
where ${\langle P_B p_2 |S| P_A p_1 \rangle}_{hard}$ is the hard IR-finite part and the IR divergent prefactor $W_{4,\text{QED}}$ can be reproduced by replacing the external particles by Wilson lines
\begin{align}
    W_{4,\text{QED}} = \bra{0} \prod_{n=1,3} \mathcal{W}_{p_n,\text{QED}}(-\infty,0) \prod_{m=2,4} \mathcal{W}_{p_m,\text{QED}}(0,\infty)\ket{0}~,
    \end{align}
    with 
    \begin{align}
    \mathcal{W}_{p,\text{QED}}(a,b)=\text{exp}\left(-i e \int_a^b ds~ p^\mu A_\mu(p s)\right)~.
\end{align}
For the case of the Coulomb background the two-point amplitude is related to the eikonal scattering from a massive particle  where $P^\mu_A=P^\mu_B=M u^\mu$ with $u$ a unit time-like vector e.g.  $u^\mu=(1,0,0,0)$. The exponential IR divergent prefactor is, using dimensional regularisation with $d=4-2\epsilon$, 
\begin{align}
\label{eq:SQED_Mass_IR}
    W_{4,QED}&=\text{exp}\Bigg\{-\frac{1}{16 \pi^2\epsilon}\sum_{n,m=A,B} \frac{e_n e_m \eta_n \eta_m}{\beta_{nm}}\left[\frac{1}{2}\ln\left( \frac{1+\beta_{nm}}{1-\beta_{nm}}\right)-i \pi \delta_{\eta_n,\eta_m}\right]\nonumber\\
 &~~~~-   \frac{1}{16\pi^2 \epsilon}\sum_{n,m=1,2} e_n e_m \eta_n \eta_m \left[\ln\left( 2|p_n\cdot p_m|\right) -i \pi \delta_{\eta_n,\eta_m}\right]\nonumber\\
 &~~~~-\frac{1}{8\pi^2 \epsilon}\sum_{\substack{n=1,2\\m=A,B}} e_n e_m \eta_n \eta_m \left[\ln \left(\frac{2|p_n\cdot P_m|}{M^2}\right) - i \pi \delta_{\eta_n,\eta_m}\right]\Bigg\}
\end{align}
where for massive particles we define the relative velocity
\begin{align}
\beta_{nm} = \left[1-\frac{ M^4}{\left(P_{n} \cdot P_{m}\right)^{2}}\right]^{1 / 2} \,. 
\end{align} 

The first line in \eqref{eq:SQED_Mass_IR} corresponds to IR divergences due to virtual photon exchange between the background particles and are not present in the two-particle amplitude where the background is entirely classical. Similarly the second line is due to photon exchange between the probe particles and is again not present in the two-particle amplitude as it is a quantum effect and would require including dynamical photons. The last term corresponds to exchange between the background and probe particles and these IR divergences are present in the classical computation of the two-point amplitude, which however only includes absorption by the probe particles and so is reduced by a factor of two.

In the case where the background particles are also massless the IR divergent prefactor can be written more compactly,
\begin{equation}
\label{eq:SQED_IR}
  W_{4,\text{QED}}=\text{exp}\Big\{-\frac{1}{8\pi^2 \epsilon}\sum_{n<m} e_n e_m \eta_n \eta_m \left[\ln\left( 2 |p_n\cdot p_m|\right) - i \pi \delta_{\eta_n,\eta_m}\right]\Big\}
\end{equation}
 where the double sum goes over all ordered pairs of legs and we have dropped $n=m$ terms in the sum which are necessary for the complete cancellation of IR divergences. These terms do not occur in the two-point amplitude and so are not relevant to our considerations. This is also consistent with the treatment of IR divergences of celestial amplitudes in \cite{Arkani-Hamed:2020gyp}. Importantly, the imaginary term only receives contributions from pairs of legs that are either both incoming or both outgoing.
 
For the Coulomb background in the charged particle rest frame we have,
\begin{align}
P_A^{\mu} \to M (1,\vec{0}) \,, \qquad P_B^{\mu}  \to M (1,\vec{0}) \,,
\label{eq:restframe-match}
\end{align} 
which implies
\begin{align}
\beta_{A A} = \beta_{B B} &= \beta_{A B} \to 0 \,, \quad P_{A} \cdot p_{m} = P_{B} \cdot p_{m}  \to -M p_m^0 \,.
\label{eq:restframe-match2}
\end{align} 
We can now focus on the mixed contributions, take the probe of charge $e_1 = e_2 = e$ and the high energy particle generating the background of charge $e_A=e_B=Q$. Then \eqref{eq:SQED_Mass_IR} becomes, since the real contributions involving $P_A$ and $P_B$ cancel out, 
\begin{align}
 W_{4,\text{QED}}^{\text{Coul}} &= \exp \left\{\frac{i e Q}{4 \pi \epsilon} \right\} \,.
 \label{eq:IR_virtual_Coul2}
 \end{align}
 
This phase is a fully classical contribution, and it is indeed what we would expect from the scattering on a stationary background within the classical eikonal approach (see \cite{Adamo:2021rfq}, for example). In particular, using the Mandelstam variables $s=-(p_1+P_A)^2$, $u=-(p_1-P_B)^2$, $t=-(p_1-p_2)^2$, the eikonal limit corresponds to $-t \ll s\simeq -u$ and in this limit the real $\ln |s|$ and $\ln|u|$ contributions cancel while the $\ln |t|$ contributions are sub-leading. The imaginary terms however do not cancel and can be identified with the IR divergent phase factors seen in the two-point amplitude in the Coulomb background \eqref{eq:QED_IR_phase}. The agreement is up to a factor of two which is due to the fact that the formula \eqref{eq:SQED_IR} includes couplings of the exchanged photons to both the background and the probe particles and so gives twice the contribution seen in the two-point amplitude \eqref{eq:QED_IR_phase}.
 
A similar calculation can be done for the electromagnetic shock wave solution by taking the massless limit for $P_A$ and $P_B$ and by taking them collinear with a null vector $q_{\text{sw}}$, $P_A=P_B=P^+ q_{\text{sw}}$. We can then use the matching condition in \eqref{eq:SQED_Mass_IR} 
\begin{align}
P_A^{\mu}  \to P^+ (1,0,0,1) \,, \qquad P_B^{\mu} \to P^+ (1,0,0,1) \,,
\label{eq:shockwave-match}
\end{align} 
and get from \eqref{eq:SQED_Mass_IR} that
\begin{align}
W_{4,\text{QED}}^{\text{sw}} &= \exp \left\{\frac{i e Q}{4 \pi \epsilon}   \right\}\,.
\label{eq:IR_virtual_Shock4}
\end{align}
This corresponds to the IR divergent phase due to photons exchanged between the massless, high energy particles of charge $e_A=e_B=Q$ generating the background and the probe particles of charge $e_1 = e_2 = e$ which is captured by the classical two-point amplitude. There is again a factor of two difference due to the two-point amplitude corresponding to the fact that the background corresponds to a single Wilson line, and therefore we are double-counting the coupling to the probe particle. 

In \cite{Arkani-Hamed:2020gyp} it was shown how to define IR finite celestial amplitudes in massless scalar QED (neglecting collinear divergences) by dressing the external charged particles by clouds of Goldstone modes. This corresponds to the Faddeev-Kulish dressing method \cite{Kulish:1970ut} but with an alternative choice of the dressing factor that is consistent with conformal invariance but not energy finiteness. The dressing of single particle states is given by
\begin{equation}
    e^{-iR^{\text QED}_k}\ket{\omega_k, z_k,\bar z_k e_k}
\end{equation}
with $R^{\text QED}_k=\eta_k e_k \Phi(z_k,\bar z_k)$ where $\Phi(z,\bar z)$ is a free, periodic, two-dimensional boson with two-point function
\begin{equation}
    \langle \Phi(z_i,\bar z_i)\Phi(z_j,\bar z_j)\rangle=\frac{1}{8\pi^2\epsilon} \ln |z_{ij}|^2
\end{equation}
and which has the interpretation as the Goldstone boson for the spontaneous breaking of the large U$(1)$ gauge symmetry \cite{Nande:2017dba}. The real part of the IR divergent prefactor of scattering amplitudes can be written as a correlation function of the operators $e^{iR^{\text{QED}}_k}$ and so is cancelled by the product of single particle dressing factors. 

To make contact with the two-point amplitude we note that properly accounting for the phase terms is essential and so we introduce two bosons $\Phi^{+}$ and $\Phi^{-}$ which have the two-point functions
\begin{equation}
    \langle \Phi^{\eta_i}(z_i,\bar z_i)\Phi^{\eta_j}(z_j,\bar z_j)\rangle=\frac{\eta_i \eta_j}{8\pi^2\epsilon} (\ln |z_{ij}|^2-i \pi \delta_{\eta_i, \eta_j} )~.
\end{equation}
We define the dressing factor for in-/out-going particles as
\begin{equation}
\label{eq:CFK_dressing}
R_k^{\mp, \text{QED}}= e_k \Phi^{\mp}(z_k,\bar z_k)
\end{equation}
where we absorb the factor of $\eta_k$ into the definition of the fields $\Phi^{\eta_k}$. To define a dressing factor for the background we 
consider the OPE of the two dressing factors for the background-generating, high-energy particles which are collinear.  As we are taking the background to be entirely classical we can take as a background dressing operator the appropriately normalised normal ordered product
\begin{equation}
    e^{iR^{\text QED}_{\text{sw}}}=\;:e^{\tfrac{i}{2}R^{-, \text{QED}}_{P_A}}e^{\tfrac{i}{2}R^{+, \text{QED}}_{P_B}}:\;
\end{equation}
so that
\begin{equation}
R^{\text QED}_{\text{sw}}=\tfrac{Q}{2} (\Phi^{+}(z_{\text{sw}},\bar z_{\text{sw}})+\Phi^{-}(z_{\text{sw}},\bar z_{\text{sw}}))~.
\end{equation}
The soft factor of the two-point function can be written as 
\begin{equation}
    \widetilde{\mathcal{A}}_2^{soft}=\langle e^{i R^{\text{QED}}_{\text{sw}}} e^{i R^{-,\text{QED}}_1} e^{i R^{+,\text{QED}}_2}\rangle~,
\end{equation}
where the terms involving the contractions between $R^{\text{QED}}_1$ and $R^{\text{QED}}_2$ are sub-leading and neglected, thus we see that the dressing factors will cancel the divergent phase and give an IR finite result. As described in \cite{Arkani-Hamed:2020gyp}, IR finite celestial amplitudes between massless scalars are obtained by dressing the conformal primary operator for outgoing or incoming states $\mathcal{O}^{\pm}_{\Delta_k}(z_k,\bar z_k)$ with $\text{exp}(-i e_k  \Phi^{\pm}(z_k,\bar z_k))$ where the dressed  operator $\hat{\mathcal{O}}^{\pm}_{\Delta_k+\delta_k}$ has a shifted dimension, with $\delta_k=-\frac{e_k^2}{8\pi^2 \epsilon}$, and is defined by appropriately accounting for the collinear singularity
\begin{equation}
  \hat{\mathcal{O}}^{\pm}_{\Delta_k+\delta_k}(z,\bar z)=\lim_{w\to z} |z-w|^{-2\delta_k}  :e^{-i e_k  \Phi^\pm(z,\bar z)}::\mathcal{O}^{\pm}_{\Delta_k}(w,\bar w):~.
\end{equation}
We similarly define a dressed shockwave operator by taking the operator product of the undressed shockwave operator $\mathcal{O}_{\text{sw}}$ and the dressing $\text{exp}(-i R^{\text{QED}}_{\text{sw}})$, 
\begin{equation}
  \hat{\mathcal{O}}_{\text{sw}}(z_\text{sw},\bar z_\text{sw})=\lim_{z\to z_\text{sw}}   :e^{-i\frac{Q}{2} (\Phi^{+}(z_\text{sw},\bar z_\text{sw})+\Phi^{-}(z_\text{sw},\bar z_\text{sw}))}::\mathcal{O}_{\text{sw}}(z,\bar z):~
\end{equation}
where, as we are treating the shockwave classically, there are no collinear singularities or shifts of dimension. The IR finite two-point amplitude in the shockwave background is then 
\begin{equation}
\widetilde{\mathcal{A}}_2^{\text{dressed}}=\langle 
\hat{\mathcal{O}}_{\text{sw}}(z_{\text{sw}},\bar z_{\text{sw}})\hat{\mathcal{O}}^{-}_{\Delta_1}(z_1,\bar z_1)
\hat{\mathcal{O}}^{+}_{\Delta_2}(z_2,\bar z_2)\rangle
\end{equation}
where the extra factors arising from the contractions between the Goldstone bosons cancel the IR divergent phases.

\subsection{Gravitational amplitudes} 
Similar results are known for the virtual infrared divergences of gravitational amplitudes \cite{Weinberg:1965nx}. Let us again denote by $p_1$, $p_3=P_A$ the incoming momenta and  $p_2$, $p_4=P_B$ the outgoing momenta. Starting with the case of scattering from a massive particle we take the momenta $p_1$ and $p_2$ as massless probe particles while $P_A$ and $P_B$ are massive background particles. The S-matrix for the four-point amplitude can again be written as
\begin{align}
\langle p_2 P_B |S|p_1 P_A \rangle = W_{4,\text{GR}} {\langle p_2 P_B |S|  p_1 P_A \rangle}_{hard} \,,
\end{align}
where ${\langle p_2 P_B |S|  p_1 P_A \rangle}_{hard}$ is the hard IR-finite part and $W_{4,\text{GR}}$ is the infrared divergent factor. The IR divergences can be again be found by replacing the external legs by Wilson lines and computing the appropriate expectation value \cite{Naculich:2011ry, White:2011yy}
\begin{align}
    W_{4,\text{GR}}=\bra{0} \prod_{n=1,3} \mathcal{W}_{p_n,\text{GR}}(-\infty,0) \prod_{m=2,4} \mathcal{W}_{p_m,\text{GR}}(0,\infty)\ket{0}~,
\end{align}
with
\begin{align}
\mathcal{W}_{p,\text{GR}}(a,b)=\text{exp}\left(-i \frac{\kappa}{2} \int_a^b ds~  p^\mu p^\nu h_{\mu\nu}(p s)\right)
\end{align}
which in $d=4-2 \epsilon$ dimensions is given by \cite{Weinberg:1965nx}
\begin{align}
W_{4,\text{GR}} &= \exp \Bigg\{\frac{G}{4 \pi \epsilon} \sum_{n,m = A,B} \eta_{n} \eta_{m} M^2 \frac{1+\beta_{nm}^{2}}{\beta_{nm} (1-\beta_{nm}^{2})^{1/2}} \left[\frac{1}{2}\ln \left(\frac{1+\beta_{nm}}{1-\beta_{nm}}\right) - i \pi \delta_{\eta_n, \eta_m} \right] \nonumber \\
&\qquad \qquad + \frac{G}{4 \pi \epsilon} \sum_{n,m=1,2} \eta_n \eta_{m} \left(2 |p_n \cdot p_m|\right) \left[\ln \left(2| p_n \cdot p_m|\right) - i \pi \delta_{\eta_n, \eta_m} \right]  \nonumber \\
&\qquad \qquad + \frac{G}{2 \pi \epsilon} \sum_{\substack{n=1,2 \\ m = A,B}} \eta_{n} \eta_{m}\left(2 |P_{n} \cdot p_{m}| \right) \left[\ln \left(\frac{2 |P_{n} \cdot p_{m}|}{M^2}\right)- i \pi \delta_{\eta_n ,\eta_m} \right] \Bigg\} \,.
\label{eq:IR_virtual}
\end{align}
This can also be derived from the one-loop soft function, by generalizing the derivation done in massless case \cite{Naculich:2011ry} to the case of massive lines \cite{Aybat:2006mz,Mitov:2010xw}. In the first two lines of \eqref{eq:IR_virtual} we can recognize the pure massive and massless contributions to the virtual IR divergences which are not reproduced by the classical two-point amplitude. The first term corresponds to including corrections to the background while the second term in \eqref{eq:IR_virtual_Schw2} is the standard massless contribution, \cite{Naculich:2011ry}, and it suppressed in classical calculations on backgrounds because it has a quantum nature. In the last line we have the mixed contributions which are present in the background two-point function and are thus captured by the classical calculation. 
When the background particles are massless we can write \eqref{eq:IR_virtual} in a compact form as in QED,
\begin{equation}
\label{eq:SGR_IR}
  W_{4,\text{GR}}=\text{exp}\Big[\frac{G}{2\pi \epsilon}\sum_{n<m} \eta_n \eta_{m} \left(2 |p_n \cdot p_m|\right) \left(\ln 2| p_n \cdot p_m| - i \pi \delta_{\eta_n, \eta_m} \right)\Big]
\end{equation}

For the Schwarzschild background, we have the same matching condition as for the Coulomb case and we can use \eqref{eq:restframe-match} and \eqref{eq:restframe-match2}. In particular, considering the mixed contribution in \eqref{eq:IR_virtual}  we get, 
\begin{align}
 W_{4,\text{GR}}^{\text{Schw}} &= \exp \left\{-\frac{i M G}{\epsilon} \left(p^0_{1}+ p^0_{2}\right)  \right\} \,.
 \label{eq:IR_virtual_Schw2}
 \end{align}
 This classical contribution can be compared with the two-point amplitude in the Schwarzschild background \eqref{eq:IR_grav} with $r_0=M$, where it can be seen, after using energy conservation $\eta_1 p_1^0 + \eta_2 p_2^0 = 0$ with $\eta_1=-\eta_2$, that there is agreement up to a factor of two which has the same physical explanation as in the QED case. A related phase has been found in the study of soft factors for low energy graviton amplitudes \cite{Laddha:2018myi}.
 
For the Aichelburg-Sexl metric we can use the matching condition in \eqref{eq:shockwave-match} which implies $P_{A} \cdot p_{j} \to -2 P^+ p_j^-$ so  we get from \eqref{eq:IR_virtual}
\begin{align}
W_{4,\text{GR}}^{\text{\text{sw}}} &= \exp \left\{-\frac{2i P^+ G}{\epsilon} \left(p^-_{1} + p^-_{2} \right)   \right\}\,,
\label{eq:IR_virtual_Shock2}
\end{align}
which is the IR divergent phase for a probe particle travelling in a gravitational shock-wave background.  Using the light-cone energy conservation  $\eta_1 p_1^- + \eta_2 p_2^- = 0$ with $\eta_1=-\eta_2$ we  again, as in all the previous cases, find a factor of two difference compared to the two-point amplitude in the Aichelburg-Sexl background \eqref{eq:IR_grav} with $r_0=P^+$ due to the identification of two Wilson lines with a single background.

As for scalar QED, it is possible to define IR finite amplitudes of massless scalars in the gravitational case by using the appropriate dressing \cite{Himwich:2020rro} which can be extended to the celestial amplitudes \cite{Arkani-Hamed:2020gyp}. 
The dressing in \cite{Himwich:2020rro} is defined in terms of a Goldstone boson $C(z, \bar{z})$ for the supertranslation symmetry such that the correlation function of exponentiated operators reproduces the IR divergences of massless scalars. In order to capture the phase information we will introduce two scalar fields $C^{\pm}(z,\bar{z})$, for incoming $(-)$ or outgoing $(+)$ particles, which have the two-point function 
\begin{align}
    \langle C^{\eta_i}(z_i, \bar{z}_i) C^{\eta_j}(z_j,\bar{z}_j)\rangle=-\frac{ \eta_i \eta_j}{4 \pi^2\epsilon}|z_{ij}|^2(\ln |z_{ij}|^2-i \pi\delta_{\eta_i,\eta_j})~.
\end{align}
 The IR divergences of massless scalars can then be reproduced by the correlation function of the exponentiated operators $e^{i R_k^{\pm,\text{GR}}}$
 \begin{align}
   W_{n, GR}=  \langle e^{i R_1^{\pm,\text{GR}}}\dots e^{i R_n^{\pm,\text{GR}}}\rangle
 \end{align}
where 
\begin{equation}
R_k^{\pm,\text{GR}}=\frac{\kappa}{2}\omega_k C^\pm(z_k, \bar{z}_k)~.
\end{equation}

We can use the same operators to reproduce the IR divergences of celestial amplitudes by making the replacement $\omega_i \to\text{exp}(\partial_{\Delta_i})$. To define IR finite two-point functions we must define a dressing for the background and, as in the QED case, this can be done by taking the collinear limit of dressings for the background particles. For the shockwave background we find the dressing
\begin{align}
    R_{\text{sw}}^{\text{GR}}=\frac{\kappa P^+}{4} (C^+(z_{\text{sw}},\bar{z}_{\text{sw}})+C^-(z_{\text{sw}},\bar{z}_{\text{sw}}))
\end{align}
so that the correlator
\begin{equation}
    \widetilde{\mathcal{M}}_2^{soft}=\langle e^{i R_{\text{sw}}^{\text{GR}}} e^{i R_1^{-,\text{GR}}}e^{i R^{+,\text{GR}}_2}\rangle
\end{equation}
correctly reproduces the IR divergences of the two-point amplitude in the shockwave background, again neglecting subleading quantum corrections. Correspondingly, we can define dressed operators 
\begin{equation}
    \hat{\mathcal{O}}^\pm_{\Delta_k}(z_k, \bar z_k)=e^{-i R^{\pm,\text{GR}}_k }\mathcal{O}^\pm_{\Delta_k}(z_k, \bar z_k)\,, \quad  \hat{\mathcal{O}}_{\Delta_{\text{sw}}}(z_{\text{sw}},\bar z_{\text{sw}})=e^{-i R^{\text{GR}}_{\text{sw}} }\mathcal{O}_{\Delta_{\text{sw}}}(z_{\text{sw}}, \bar z_{\text{sw}})\,,
\end{equation}
such that the 
the IR finite two-point amplitude is
\begin{equation}
\widetilde{\mathcal{M}}_2^{\text{dressed}}=\langle 
\hat{\mathcal{O}}_{\text{sw}}(z_{\text{sw}}, \bar z_{\text{sw}})\hat{\mathcal{O}}^{-}_{\Delta_1}(z_1, \bar z_1)
\hat{\mathcal{O}}^{+}_{\Delta_2}(z_2, \bar z_2)\rangle~.
\end{equation}
The extension of the Goldstone boson dressing to massive particles, which correspond to the Schwarzschild background, has been discussed in \cite{Himwich:2020rro} however it is slightly more involved and as the two-point function in massive backgrounds does not yet have such a clear interpretation as a three-point function we leave this case to future work.

\section{Conclusions}

Flat space scattering amplitudes have been recast as CFT correlators on the celestial sphere, in a way that makes manifest the symmetries we would expect for an underlying holographically dual theory. The asymptotic symmetry analysis, however, suggests the existence of a celestial conformal field theory also for more general asymptotically flat backgrounds. The main goal of this paper was to start exploring its properties for Kerr-Schild backgrounds in four spacetime dimensions. 

To that end we focused on a particularly simple object: the two-point amplitude for the scattering of a massless field on top of such backgrounds which we have computed using the method of Boulware and Brown. For general asymptotically flat curved spacetimes the global Lorentz symmetry of the S-matrix is broken, but this does not prevent us from giving an interpretation for the scattering of conformal primaries in these geometries. Indeed, at least for point-like backgrounds all the semiclassical two-point amplitudes can be related to suitable Lorentz-invariant four-point amplitudes in flat space which do have a clear CCFT interpretation. Moreover, we can define classical point-like backgrounds as the ones generated by three-point amplitudes with an off-shell coherent emission in the $\hbar \to 0$ limit \cite{Monteiro:2020plf} which guarantees that a proper holographic description must exists for such backgrounds, thanks to the holographic map from S-matrix elements to CCFT correlators. 
The backgrounds in this class for which we computed celestial amplitudes are Coulomb, Schwarzschild and the electromagnetic and gravitational shockwave. We have also studied the classically spinning counterparts of these solutions, i.e. $\sqrt{\text{Kerr}}$, $\text{Kerr}$ and their corresponding ultra-boosted limits, which exhibit interesting finite-size spin effects and might ultimately require an intrinsic worldsheet-type description.

We find that the celestial two-point amplitudes on the electromagnetic and the gravitational shockwave backgrounds at leading order in the coupling exhibit the structure of vanilla three-point functions in a CFT. This can be attributed to the fact that such backgrounds can be interpreted directly as conformal primaries on the celestial sphere \cite{Pasterski:2020pdk}. Interestingly, the kinematic delta functions of flat space three-point amplitudes are absent, suggesting that no shadow or light-ray transform prescription~\cite{Fan:2021isc,Crawley:2021ivb,Sharma:2021gcz,De:2022gjn} that alleviates them in flat space is needed. 
For massive scalar point-like backgrounds like Coulomb and Schwarzschild, we similarly find a power-law, rather than distributional, behaviour in the celestial sphere coordinates for the two-point function which calls for an interpretation of the corresponding bulk solutions in terms of dual operators in the CCFT. We also considered classically spinning backgrounds, both massless and massive, where new interesting features arise. For example, the celestial two-point amplitudes for the $\sqrt{\text{Kerr}}$, Kerr and spinning shockwave solutions are well-behaved in the UV and not restricted to the principal series provided an appropriate analytic continuation in spin is performed. All these features make the study of these amplitudes very interesting from the holographic perspective, and we are looking forward to finding a full description of such bulk solutions in terms of conformal primary operators. 

In AdS/CFT, the on-shell action plays a special role in the derivation of the holographic two-point function as it localizes along the boundary. A similar calculation can be done for asymptotically flat backgrounds: we have shown that the boundary on-shell action evaluated at null infinity is the effective source for the wave equation localized on the celestial sphere, both for spinless and spinning wave scattering. The boundary action is therefore the generating functional for tree-level correlation functions on the celestial CFT, consistent with the holographic principle.

Finally, we have studied the infrared sector for wave scattering on top of backgrounds. As expected, infrared divergences exponentiate into a divergent phase dressing for the two-point amplitude. While this is irrelevant for cross-section observables, it is crucial in order to define an infrared finite S-matrix. For point-like backgrounds we have derived a conformal Faddeev-Kulish dressing along the lines of \cite{Arkani-Hamed:2020gyp}, which removes the IR phase divergences at all orders in perturbation theory and which is generated by a pair of Goldstone bosons arising from the spontaneous breaking of large gauge and gravitational symmetries. This solves the problem of describing the infrared dynamics for two-point amplitudes on asymptotically flat point-like backgrounds. We expect our result to be valid also for spinning backgrounds, since the spin dependence drops out for the infrared dynamics as we showed explicitly by considering the conformally soft limit of wave scattering on such backgrounds.

We conclude with some open questions.  An important step is to understand what operators or combination thereof represents bulk massive background solutions, beyond the simple shockwave case. It is conceivable that a suitable combination of conformal primaries can represent physical massive black holes in the CCFT, and this is a promising direction for further studies. Here, we have only considered in detail the scattering of massless scalars. The perturbiner method, however, can also be extended to the scattering of spinning waves, as we showed in the appendices, and it would be interesting to study spinning two-point amplitudes.  Finally, it would be nice to understand to what extent the energy scale of bulk gravitational solutions is going to affect the dual description in the CCFT. So far we have treated such energy scale as a parameter by considering the scattering of external conformal primaries on the original geometry, but perhaps there exists an alternative intrinsic dual description of geometry itself purely in terms of conformal primaries. A last tantalizing possibility would be to study the two-point function with ingoing boundary conditions at the horizon for asymptotically flat black hole solutions, which would allow the exploration of the structure of the dual CFT description.  

\section*{Acknowledgements}
We would like to thank Tim Adamo, Davide Billo, Andrea Cristofoli, Alok Laddha and Nathan Moynihan for interesting discussions.
TMCL was supported by Science Foundation Ireland through grant 15/CDA/3472. AP is supported by the European Research Council (ERC) under the European Union’s Horizon 2020 research and innovation programme (grant agreement No 852386). RG has received funding from the European Union's Horizon 2020 research and innovation program under the Marie Sklodowska-Curie grant agreement No.764850 ``SAGEX''.

\appendix{

\section{Spinning wave scattering on backgrounds}\label{app:spinningwavescattering}

We generalize the analysis of scalar wave scattering on gravitational backgrounds, section~\ref{sec:scalarwavescattering}, to the spinning case with $s=1,2$. 

Let's consider first a $U(1)$ vector field $A_{\mu}$ minimally coupled to a gravitational Kerr-Schild background, where the corresponding action is
\begin{align}
\mathcal{S} = - \frac{1}{4} \int d^4 x\, \sqrt{-g} g^{\mu \rho} g^{\nu \sigma} F_{\mu \nu} F_{\rho \sigma} + \int d^4 x\, \sqrt{-g} J^{\mu} A_{\mu} \,.
\label{eq:action_spin1}
\end{align}
We have the following equation of motion
\begin{align}
\underline{\DAlambert} A^{\mu} - \underline{\nabla}^{\mu} \underline{\nabla}_{\nu} A^{\nu} - \underline{R}^{\mu}_{\,\,\nu} A^{\nu} = -J^{\mu}\, , 
\label{eq:spin1_pert}
\end{align}
where the underlined quantities refer to the background Kerr-Schild metric. To construct the two-point amplitude with the perturbiner method, it is convenient to work with cartesian Kerr-Schild coordinates. The equation of motion in Lorenz gauge can be written in terms of an effective source current
\begin{align}
\eta^{\alpha \beta} \partial_{\alpha} \partial_{\beta} A^{\mu} &= J^{\mu}_{\text{eff}} \,,\qquad   \nabla_{\mu} A^{\mu} = 0 \,, 
\end{align}
where 
\begin{align}
J^{\mu}_{\text{eff}} &= \partial_{\alpha} \left(h^{\alpha \beta} \underline{\nabla}_{\beta} A^{\mu} \right) -\eta^{\alpha \beta} \partial_{\alpha} (\underline{\Gamma}^{\mu}_{\beta \gamma} A^{\gamma}) - \underline{\Gamma}^{\mu}_{\lambda \sigma} g^{\lambda \beta } \underline{\nabla}_{\beta} A^{\sigma}  + \underline{R}^{\nu}_{\,\,\mu} A^{\mu} - J^{\mu}\,.
\label{eq:spin1_sol}
\end{align}
This can be inverted to give the classical solution in terms of the source
\begin{align}
A^{\mu}_{\text{cl}}(x) &= \int \frac{d^4 p}{(2 \pi)^4} \frac{e^{i p \cdot x}}{(p^0)^2 - |\vec{p}|^2 + i \epsilon} \bar{J}^{\mu}_{\text{eff}}(p^0,\vec{p})\,
\label{eq:A_fullsol}
\end{align}
which can in practice be solved iteratively for weakly curved backgrounds. 
We can perform the standard LSZ reduction to define the two-point amplitude for spin 1 fields on the Kerr-Schild gravitational background
\begin{align}
\mathcal A_2(p_1^{\sigma_1},p_2^{\sigma_2})&=-\lim_{p_1^2\to 0}\lim_{p_2^2\to 0} p_1^2p_2^2 \,  \varepsilon^{*\sigma_1}_{\mu}(-p_1) \varepsilon^{\sigma_2}_{\nu}(p_2) \frac{\delta \bar A^{\mu}_{\text{cl}}(-p_1)}{\delta \bar J_{\nu}(p_2)} \,.
\end{align}
A similar procedure can be carried out for graviton perturbations on a Kerr-Schild background, by using the Einstein-Hilbert Lagrangian coupled to a matter source
\begin{align}
\mathcal{S} &= \frac{2}{\kappa^2} \int d^4 x\, \sqrt{-G} R[G] + \int d^4 x\, \sqrt{-G} \mathcal{L}_{\text{matter}} \,, \qquad J_{\mu \nu} \equiv -\frac{1}{\sqrt{-G}} \frac{\partial (\sqrt{-G} \mathcal{L}_{\text{matter}} )}{\partial G^{\mu \nu}} \, .
\end{align}
We can then express the total metric as the sum of the spin-2 tensor perturbation $H_{\mu \nu}$ and the Kerr-Schild background $g_{\mu \nu}$
\begin{align}
G_{\mu \nu} = g_{\mu \nu} + H_{\mu \nu} \,, 
\end{align}
and we can express the perturbative wave equation in terms of trace-reversed perturbations
\begin{align}
H_{\mu \nu}^{\text{tr}} &\equiv H_{\mu \nu} - \frac{1}{2} g_{\mu \nu} H^{\alpha}_{\,\,\alpha}
 \,, \qquad   \nabla_{\mu} (H^{\text{tr}})^{\mu \nu} = 0 \,. 
\end{align}
The wave equation for the spin 2 perturbation on the Kerr-Schild background is therefore defined as
\begin{align}
\underline{\DAlambert} (H^{\text{tr}})^{\mu \nu}   - \underline{R}_{\alpha \beta} \left(g^{\mu \nu} (H^{\text{tr}})^{\alpha \beta}-(H^{\text{tr}})^{\mu \nu} g^{\alpha \beta}+ 2 g^{\beta(\mu} (H^{\text{tr}})^{\nu) \alpha}\right)& \nonumber\\
&\, \kern-120pt+ 2 \underline{R}^{\mu \,\,\,\, \nu}_{\,\,\,\alpha \,\,\,\, \beta} (H^{\text{tr}})^{\alpha \beta}=2 \,\delta E^{\mu \nu}[H^{\text{tr}}] - J^{\mu \nu}(x)\,, 
\label{eq:spin2_pert}
\end{align}
where $\delta E_{\mu \nu}$ includes all the quadratic, cubic and higher order contributions of $(H^{\text{tr}})_{\mu \nu}$ to the Einstein tensor and the equation has to be solved self-consistently in powers of $\kappa$. At each order in perturbation theory, an effective source can be defined for $(H^{\text{tr}})^{\mu \nu}$
\begin{align}
J^{\mu \nu}_{\text{eff}}(x) &\equiv \partial_{\alpha} \left(h^{\alpha \beta} \underline{\nabla}_{\beta} (H^{\text{tr}})^{\mu \nu} \right) -2 \eta^{\alpha \beta} \partial_{\alpha} (\underline{\Gamma}^{(\mu}_{\beta \rho} (H^{\text{tr}})^{\nu) \rho}) - 2 g^{\alpha \beta} \underline{\Gamma}^{(\mu}_{\alpha \rho}\underline{\nabla}_{\beta} (H^{\text{tr}})^{\nu) \rho}\nonumber \\ 
&\qquad  - 2 \underline{R}^{\mu \,\,\,\, \nu}_{\,\,\,\alpha \,\,\,\, \beta} (H^{\text{tr}})^{\alpha \beta} + \underline{R}_{\alpha \beta} \left[g^{\mu \nu} (H^{\text{tr}})^{\alpha \beta}-(H^{\text{tr}})^{\mu \nu} g^{\alpha \beta}+2 g^{\beta(\mu} (H^{\text{tr}})^{\nu) \alpha}\right] \nonumber \\
&\qquad + 2 \,\delta E^{\mu \nu}[H^{\text{tr}}]- J^{\mu \nu}(x)
\label{eq:spin2_sol}
\end{align}
such that the equation of motion is 
\begin{equation}
\eta^{\alpha \beta} \partial_{\alpha} \partial_{\beta} (H^{\text{tr}})^{\mu \nu} = J^{\mu \nu}_{\text{eff}} \,
\end{equation}
which can be solved iteratively for $(\bar{H}^{\text{tr}})^{\mu \nu}_{cl}$ as a function of the source.
As before we can perform the standard LSZ reduction to define the perturbative two-point amplitude 
\begin{align}
\mathcal M_2(p_1^{\sigma_1},p_2^{\sigma_2})&=-\lim_{p_1^2\to 0}\lim_{p_2^2\to 0} p_1^2p_2^2 \,  \varepsilon^{*\sigma_1}_{\mu \nu}(-p_1) \varepsilon^{\sigma_2}_{\alpha \beta}(p_2) \frac{\delta (\bar{H}^{\text{tr}})^{\mu \nu}_{\text{cl}}(-p_1)}{\delta \bar J_{\alpha \beta}(p_2)} \,.
\end{align}
Summarizing, it is always possible to define an effective source for the wave scattering of massless spin-$s$ perturbations on Kerr-Schild backgrounds in perturbation theory. If we collectively define our scalar, vector and tensor perturbations as
\begin{align}
\Psi_I = \{\phi, A_{\mu}, H^{\text{tr}}_{\mu \nu}\} \,,
\end{align}
then the solution wave equation in cartesian Kerr-Schild coordinates can be always written in the form
\begin{align}
\Psi_{I,\text{cl}}(x) &= \int \frac{d^4 p}{(2 \pi)^4} \frac{e^{i p \cdot x}}{(p^0)^2 - |\vec{p}|^2 + i \epsilon} \bar{J}_{I,\text{eff}}(p^0,\vec{p}) \, .
\label{eq:waveeq_general}
\end{align}

\section{Spinning boundary on-shell action localization}\label{app:spinningbdyaction}

We show here that we can extend the relation between the boundary on-shell action and the generating functional of the two-point function for a $U(1)$ gauge field minimally coupled to a Kerr-Schild gravitational background, working for simplicity in radial gauge $A_r = 0$. 

The original action in \eqref{eq:action_spin1} is equivalent to
\begin{align}
\mathcal{S}&=\frac{1}{2} \int \mathrm{~d}^{4} x \sqrt{-g} A^{\mu} \left(\delta_{\mu}^{\nu} \underline{\nabla}^{\rho} \underline{\nabla}_{\rho} -\underline{\nabla}_{\mu} \underline{\nabla}^{\nu} - \underline{R}_{\mu}^{\,\,\nu}\right) A_{\nu}\nonumber \\
&-  \int \mathrm{~d}^{4} x \sqrt{-g} g^{\mu \rho} g^{\nu \xi} \underline{\nabla}_{[\mu} \left(A_{\nu]} (\underline{\nabla}_{[\rho} A_{\xi]}) \right)  \,,
\end{align}
where the boundary term can be now isolated 
\begin{align}
\mathcal{S}_{\text{bdy}}&= - \sum_{\eta=\pm} \int_{\mathscr{I}^{\eta}} \mathrm{d}^{3} x  \sqrt{-g_{\text{ind}}}  A^{[\xi}(x) n^{\rho]} \underline{\nabla}_{[\rho} A_{\xi]}(x) \,.
\end{align}
We can now write the incoming and the outgoing gauge theory solutions as
\begin{align}
A^{\mu}_{\text{in},\sigma}(x) &= \varepsilon^{*\mu}_{\sigma}(p) e^{i p \cdot x} \,,  \nonumber \\ A^{\mu}_{\text{out}}(t \mp r,r,\hat{x}) &= -c_{\pm} \frac{1}{4 \pi r} \int_{\mathbb{R}} \frac{d \omega_k}{2 \pi} \,e^{\mp i \omega_k (t \mp r)} \bar{J}^{\mu}_{\text{eff}}(\pm \omega_k, \omega_k \hat{x}) + \mathcal{O}(r^{-2}) \,,
\end{align}
where we have considered only the leading radiative component of the gauge field around the null boundaries $\mathscr{I}^{\pm}$. This means that we can write the in/out boundary contribution as a function of the momentum for the future null boundary as
\begin{align}
\mathcal{S}^{\text{in/out}}_{\mathscr{I}^+}(p^{\sigma}) =& -\frac{1}{(8 \pi^2)} c_{+} \lim_{r \to \infty}  \, r  \int_{\mathscr{I}^{+}} d u d \theta d \varphi \, \sin(\theta)\,  \nonumber \\
& \times  \int_{\mathbb{R}} d \omega_k \Bigg[e^{-i p \cdot x} \varepsilon_{\sigma}^{*[\xi}(p) n^{\rho]}  \,\underline{\nabla}_{[\rho} \left(e^{-i \omega_k u} \bar{J}_{\text{eff},\xi]}( \omega_k,\omega_k \hat{x}) \right) \nonumber \\
&\qquad \qquad \qquad- \varepsilon_{\sigma,\xi}^{*}(p)  \left(e^{-i \omega_k u} \bar{J}_{\text{eff}}^{[\xi}(\omega_k,\omega_k \hat{x}) n^{\rho]} \right) \underline{\nabla}_{\rho} e^{i p \cdot x} \Bigg] \,,
\end{align}
and for the past null boundary
\begin{align}
\mathcal{S}^{\text{in/out}}_{\mathscr{I}^-}(p^{\sigma}) =& -\frac{1}{(8 \pi^2)} c_{-} \lim_{r \to \infty} \,r  \int_{\mathscr{I}^{-}} d v d \theta d \varphi \, \sin(\theta)\,  \nonumber \\
& \times  \int_{\mathbb{R}} d \omega_k \Bigg[e^{i p \cdot x} \varepsilon_{\sigma}^{*[\xi}(p) n^{\rho]}  \,\underline{\nabla}_{[\rho} \left(e^{i \omega_k v} \bar{J}_{\text{eff},\xi]}(-\omega_k,\omega_k \hat{x}) \right) \nonumber \\
&\qquad \qquad\qquad \qquad- \varepsilon_{\sigma,\xi}^{*}(p)  \left(e^{i \omega_k v} \bar{J}_{\text{eff}}^{[\xi}(-\omega_k,\omega_k \hat{x}) n^{\rho]} \right) \underline{\nabla}_{\rho} e^{i p \cdot x} \Bigg] \,.
\end{align}
We notice that in radial gauge $A_{\text{in}}^r = J_{\text{eff},r} = 0$ some of the terms are vanishing\footnote{It is an interesting problem to extend this derivation to other gauges, like the harmonic one.}, and following some steps similar to the scalar case, we are left with
\begin{align}
\mathcal{S}^{\text{in/out}}_{\mathscr{I}^-\cup \mathscr{I}^+}(p^{\sigma}) &= \left(\frac{c_+ + c_-}{2}\right) \varepsilon_{\sigma}^{*\xi}(p)  \bar{J}_{\text{eff},\xi}(\omega_{p},\omega_{p} \hat{p}) \,.
\label{eq:boundary_onshell_spin1}
\end{align}
which depends on the contour prescription as in section \ref{sec:bdyaction}. Therefore, we can conclude that the boundary on-shell spinning action is the generating functional of the two-point amplitude:
\begin{align}
\mathcal M_2(p_1^{\sigma_1},p_2^{\sigma_2})&=-\lim_{p_1^2\to 0}\lim_{p_2^2\to 0} p_1^2p_2^2  \varepsilon^{*\sigma_1}_{\mu}(-p_1) \varepsilon^{\sigma_2}_{\nu}(p_2) \frac{\delta\bar A^{\mu}_{\text{out}}(-p_1)}{\delta \bar J_{\nu}(p_2)} \nonumber \\
&= \lim_{p_1^2\to 0}\lim_{p_2^2\to 0} p_2^2 \, \varepsilon^{*\sigma_1}_{\mu}(-p_1) \varepsilon^{\sigma_2}_{\nu}(p_2) \frac{\delta \bar{J}^{\mu}_{\text{eff}}(-p_1)}{\delta \bar J_{\nu}(p_2)} \nonumber \\
&= \lim_{p_1^2\to 0}\lim_{p_2^2\to 0} p_2^2 \, \varepsilon^{\sigma_2}_{\nu}(p_2) \frac{\delta \left[\left(\frac{2}{c_+ + c_-}\right) \mathcal{S}^{\text{in/out}}_{\mathscr{I}^-\cup\mathscr{I}^+}(-p_1^{\sigma_1})\right]}{\delta \bar J_{\nu}(p_2)} \,,
\end{align}
This calculation can be generalized also to the tensor source of graviton perturbations in \eqref{eq:spin2_sol}, but the steps are quite involved we leave this for future work.  It has been shown explicitly in \cite{Vitenti:2012cx} how to isolate the boundary terms in the Einstein-Hilbert action up to quadratic order in the graviton perturbation, and we believe that the procedure can then be generalized recursively also at higher orders in perturbation theory.

\section{Hyperbolic slicings of Schwarzschild and Kerr}\label{app:slicing}

We have shown in section \ref{sec:celestialwavescattering} that the wave scattering on asymptotically flat backgrounds can be recast as a celestial two-point function on the CCFT.  A complementary approach to celestial holography in Minkowski spacetime makes use of the so-called hyperbolic slicing, as shown in the seminal work of Soludukhin and de Boer \cite{deBoer:2003vf}. The relevance is two-fold. First of all, this allows to make contact with the standard AdS$_3$/CFT$_2$ holography \cite{Ball:2019atb,Cheung:2016iub}. Second, this gives a manifestly Lorentz-invariant description for the bulk physics in a way that smoothly connects to the celestial conformal field theory (indeed Weyl-invariant theories are very interesting models for flat space holography\footnote{This coordinate system was first studied by Dirac in 1949 for the so-called point-form quantization \cite{Dirac:1949cp}, where the dilatation operator is the time-translation Hamiltonian in the bulk. Then this was revisited further in the early 80's by a variety of people \cite{Fubini:1972mf,Macdowell:1972ef,Gromes:1974yu,Sommerfield:1974fa,diSessa:1974ve,DiSessa:1974xd} in order to understand dual resonance models at the early stages of the development of string theory.}).

Here we show that we can also provide, at least locally in some coordinate patch, a hyperbolic slicing of asymptotically flat space solutions like Schwarzschild and Kerr in 4 dimensions starting from their Kerr-Schild form. Besides the importance of showing how the Euclidean AdS$_3$ geometry and its conformal boundary arises from these geometries in the future and past Milne regions, this justifies the use of the conformal flat space basis for the scattering on top of those backgrounds. Moreover, it might also be relevant for setting up the holographic bulk reconstruction method in asymptotically flat spacetimes, which is well-developed in the AdS/CFT literature \cite{deHaro:2000vlm,Skenderis:2002wp}. 

\subsection{Warm up: Minkowski flat metric}

Let's start with the usual Minkowski spacetime in cartesian coordinates $(x^0=t,x^1,x^2,x^3)$
\begin{align}
d s_{\text {Mink}}^{2}=\eta_{\mu \nu} d x^{\mu} d x^{\nu}\, .
\end{align}
In this case the change of coordinates we are seeking boils down to choosing Minkowski proper distance as the new time coordinate $\tau = \frac{1}{2} \log(-x^2)$. In detail \cite{Cheung:2016iub}
\begin{align}
x^{\mu} &= \left( \frac{1}{2} \rho e^{\tau} \left(\frac{1 + |z|^2}{\rho ^2}+1\right),\frac{(z + \bar{z})}{2 \rho} e^{\tau},-\frac{i (z-\bar{z})}{2 \rho} e^{\tau},\frac{1}{2} \rho e^{\tau} \left(1-\frac{1-|z|^2}{\rho^2}\right) \right) \, ,
\label{eqn:coord_cart_to_new}
\end{align}
where we have expressed our Cartesian coordinates in terms of  $(\tau,\rho,z,\bar{z})$. Clearly $(t,x^1,x^2,x^3) \in \mathbb{R}^4$, while $(\tau,\rho,z,\bar{z}) \in \mathbb{R}^2 \times \mathbb{C}^2$. If we express everything in terms of spherical coordinates $(t,r,\theta,\varphi) \in  \mathbb{R} \times (0,+\infty) \times [0,\pi) \times [0,2\pi)$ we get
\begin{align}
t &= \frac{1}{2} \rho e^{\tau} \left(\frac{1+|z|^2}{\rho ^2}+1\right)  \, , \qquad \,\,\,\, r = \frac{1}{2} e^{\tau} \mathcal{G}(z,\bar{z},\rho)  \, ,\nonumber \\
\theta &= \arccos\left(\frac{\rho^2+|z|^2-1}{\rho \,\mathcal{G}(z,\bar{z},\rho)}\right) \, ,\qquad \varphi = i \arctanh\left(\frac{\bar{z}-z}{z+\bar{z}}\right)  \, ,
\label{eqn:coord_rad_to_new}
\end{align}
where we have conveniently defined a function 
\begin{align}
\mathcal{G}(z,\bar{z},\rho) = \sqrt{\frac{\left((\rho -1)^2 + |z|^2\right) \left((\rho +1)^2 + |z|^2\right)}{\rho^2}}
\label{eqn:def_G}
\end{align}
which has a singularity on the light-cone. By applying now the change of coordinates \eqref{eqn:coord_rad_to_new} to the Minkowski metric we get
\begin{align}
d s^2_{\text{Mink}}=& e^{2 \tau}\left(-d \tau^{2}+d s_{\text{AdS}_{3}}^{2}\right) \, , \qquad d s_{\text{AdS}_{3}}^{2} = \frac{1}{\rho^{2}}\left(d \rho^{2}+d z d \bar{z}\right)\, .
\end{align}
%

\subsection{Schwarzschild metric}

We use the Kerr-Schild form of the Schwarzschild metric
\begin{align}
d s^2_{\text{Schw},(\pm)}= d s_{\text{Mink}}^2 + \frac{2 G M}{r} \left(d t^{\prime} \pm d r \right)^2\, ,
\label{eq:SchwKS}
\end{align}
where the sign $\pm$ denotes the two possible choices for this coordinate system, which is well-adapted to ingoing ($+$ sign) or outgoing ($-$ sign) null geodesics in the Schwarzschild geometry. Indeed, \eqref{eq:SchwKS} is usually called Eddington-Finkelstein system of coordinates\footnote{This follows from defining a new the time coordinate $t^{\prime}$ from the standard Schwarzschild representation
\begin{align}
d t^{\prime} = d t \pm \frac{d r}{1 - 2 G M/r}\, ,
\end{align}
where $t$ is the coordinate time.
}.

This representation of the metric is highly convenient for the holographic approach: the principal ingoing/outgoing null directions provide an intrinsic characterization of the spacetime in the dual picture. In this case we would like to use the Schwarzschild proper distance as our new coordinate time: in EF coordinates, this will approach again $\tau \to \frac{1}{2} \log(-x^2)$ at large distances like in flat space. We can therefore set
\begin{align}
t^{\prime} &= \frac{1}{2} \rho e^{\sigma} \left(\frac{\zeta \bar{\zeta}+1}{\varrho ^2}+1\right)\, , \qquad \,\,\,\, r = \frac{1}{2} e^{\sigma} \mathcal{G}(\zeta,\bar{\zeta},\varrho)\, , \nonumber \\
\theta &= \arccos\left(\frac{\varrho^2+\zeta \bar{\zeta}-1}{\varrho \mathcal{G}(\zeta,\bar{\zeta},\varrho)}\right)\, ,\qquad \varphi = i \arctanh\left(\frac{\bar{\zeta}-\zeta}{\zeta+\bar{\zeta}}\right) \, ,
\label{eqn:hyper_coord}
\end{align}
where $(\sigma,\varrho,\zeta,\bar{\zeta})$ are the analogue of $(\tau,\rho,z,\bar{z})$ for equation \eqref{eqn:coord_rad_to_new} with the crucial replacement $t \to t^{\prime}$. We then obtain, focusing for simplicity to outgoing null geodesics, that for each slice labelled by the new time coordinate $\sigma$ the metric becomes asymptotically  
\begin{align}
d s^2_{\text{Schw},(-)} \Big|_{d \sigma = 0} &= e^{2 \sigma} \left[\frac{(d \varrho)^2 + d\zeta d\bar{\zeta}}{\varrho^2}\right] \nonumber \\
& +  \varrho e^{\sigma} \frac{4 G M}{(1+|\zeta|^2)^3} (d \varrho)^2 -\varrho^2 e^{\sigma}\frac{8 G M d\varrho  (\bar{\zeta} d \zeta + \zeta d\bar{\zeta})}{(1+|\zeta|^2)^4}  + \mathcal{O}\left(\varrho^3\right)
\label{eq:Schw-hyperbslic}
\end{align}
in the expansion of $\varrho$ near the boundary (identified with $\varrho =0$). We see that we find exactly an empty AdS$_3$ metric up to order $\varrho^0$. The Ricci scalar on each slice up to order $\varrho^6$ reads
\begin{align}
R_{\sigma-\text{slice},(-)} =& -6 e^{-2 \sigma} + \mathcal{O}(\varrho^5) \,,
\end{align}
which confirms the interpretation of the AdS$_3$ for the asymptotic structure on each slice\footnote{This shows also that further coordinate transformations can be done to achieve an AdS$_3$-like metric structure up to higher orders in the $\varrho$ variable on each hyperbolic slice.}. A completely parallel calculation can be done for ingoing null geodesics $(+)$: we have compared the two cases in Fig.\ref{fig:hyperbolic_slicing}, where the region of validity for this system of coordinates is also highlighted.

\begin{figure}[!htb]
    \begin{subfigure}{0.5\textwidth}
        \centering
        \includegraphics[scale=0.80]{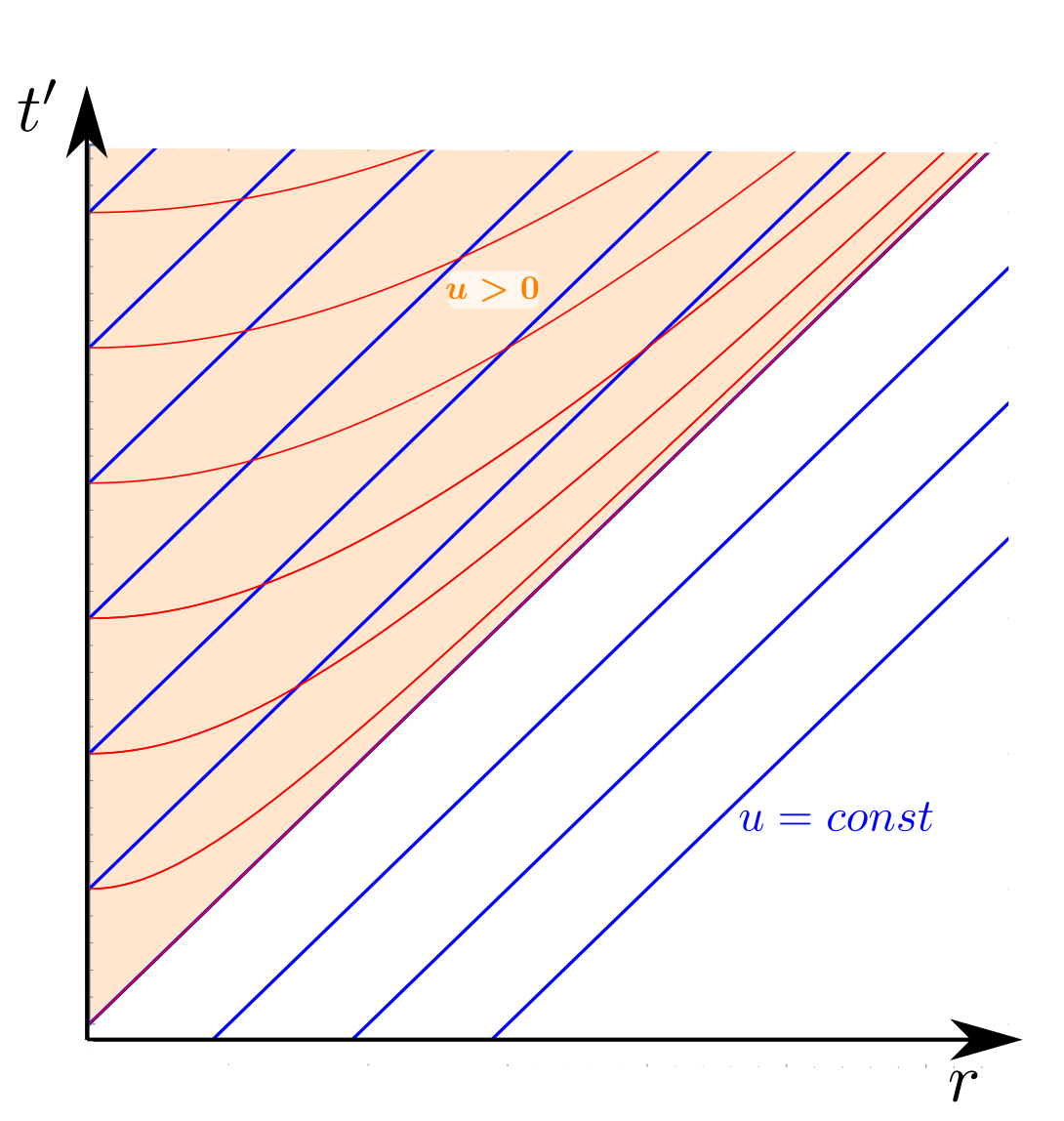}
        \caption{Outgoing case: $u = t' - r$}
        \label{fig:outgoing}
    \end{subfigure}
    \begin{subfigure}{0.5\textwidth}
        \centering
        \includegraphics[scale=0.80]{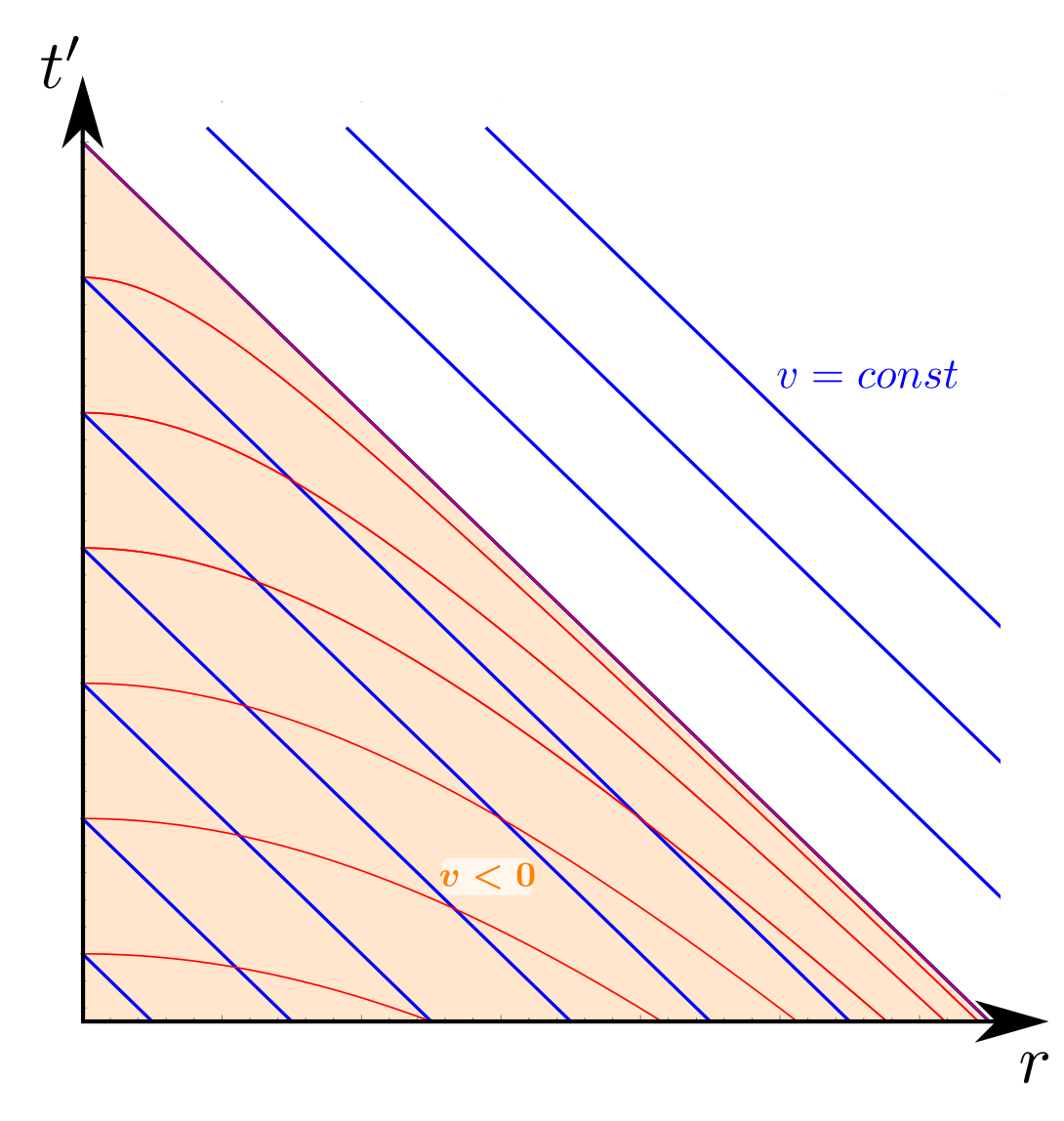}
        \caption{Ingoing case: $v = t' + r$}
        \label{fig:ingoing}
    \end{subfigure}
\caption{The region in orange shows, in the $(r,t')$ space, the patch covered by the hyperbolic system of coordinates adapted to outgoing (see Fig.\ref{fig:outgoing}) or ingoing (see Fig.\ref{fig:ingoing}) null geodesics. The straight lines in blue represent null geodesics, while the red lines represent (one side of) the hyperbolic slices at fixed time $\sigma$.}
\label{fig:hyperbolic_slicing}
\end{figure}

\subsection{Kerr metric}
As for Schwarzschild, it is highly convenient to use the Kerr-Schild form of the Kerr metric (see for example \cite{Poisson:2009pwt,Frolov:1998wf})
\begin{align}
d s^2_{\text{Kerr},(\pm)} = d s^2_{\text{Mink}} + \frac{2 G M \tilde{r}}{a^2 \cos^2\theta +\tilde{r}^2} \left(dr \pm dt - a \sin ^2\theta d\varphi \right)^2 \, ,
\label{eq:Kerr_KS}
\end{align}
where $\tilde{r}$ is defined implicitly from 
\begin{align}
\tilde{r}^4 - \tilde{r}^2 \left(r^2-a^2\right)-a^2 (r \cos\theta)^2 = 0 .
\end{align}
It is worth noticing that $\tilde{r} \to r$ for $a \to 0$: the Kerr metric in the Kerr-Schild form gives manifestly the Schwarzschild metric in the Kerr-Schild form when $a \to 0$.

Let us focus on the coordinates adapted to outgoing null geodesics. If apply the transformation \eqref{eqn:hyper_coord} to the Kerr metric, on each hyperbolic slice we get, in the small $\varrho$ expansion,
\begin{align}
d s^2_{\text{Kerr},(-)} \Big|_{d \sigma = 0} &= e^{2 \sigma} \left[\frac{(d \varrho)^2 + d\zeta d\bar{\zeta}}{\varrho^2}\right]  \nonumber \\
&+ \varrho  e^{\sigma } \frac{4 G M (d \varrho)^2}{(1+|\zeta|^2)^3} + \varrho a\frac{16 G M i d \varrho (\zeta d\bar{\zeta}-\bar{\zeta} d\zeta)}{(1+|\zeta|^2)^4} -\varrho e^{-\sigma } a^2\frac{16 G M (\bar{\zeta} d\zeta-\zeta d\bar{\zeta})^2}{(1+|\zeta|^2)^5}\nonumber \\
& + \varrho^2 a \frac{16 i G M \left(\bar{\zeta}^2 (d\zeta)^2-\zeta^2 (d\bar{\zeta})^2\right)}{(1+|\zeta|^2)^5} -\varrho^2 e^{\sigma } \frac{8 G M d \varrho (\bar{\zeta} d\zeta+\zeta d\bar{\zeta})}{(1+|\zeta|^2)^4} + \mathcal{O}(\varrho^3)
\label{eq:Kerr-hyperbslic}
\end{align}
which again confirms the interpretation of AdS$_3$ plus corrections in the $\varrho \to 0$ expansion. Moreover, if we compute the Ricci scalar on each $\sigma$-slice we find, as expected,
\begin{align}
R_{\sigma-\text{slice},(-)} = -6 e^{-2 \sigma}+ \mathcal{O}(\varrho^2).
\end{align}
As for Schwarzschild, the procedure can be repeated for ingoing null geodesics and the region of validity of this construction is highlighted again in Fig.\ref{fig:hyperbolic_slicing}.

\subsection{Gaussian normal coordinates for Schwarzschild and Kerr}
A Gaussian normal coordinate system provides a foliation of spacetime with spacelike hypersurfaces, in way that is convenient for the holographic interpretation (see \cite{deBoer:2003vf}). We would like the new $\sigma$ time coordinate to measure the proper time of "stationary" observers, i.e. observers with constant spatial coordinates. These coordinates always exist, at least locally:  one merely takes each spatial position on the 3-manifold of the foliation $\Sigma_{\sigma}$, builds the geodesic orthogonal to $\Sigma_{\sigma}$ and takes these to define the new proper time\footnote{The related observers are called fundamental comoving observers, and the fundamental property of this system is that geodesics remain orthogonal to $\Sigma_{\sigma}$ for all $\sigma$ in the region of validity of the construction.}. 

Instead of explicitly solving the geodesic equation (which would be quite cumbersome in our coordinates), we will perform a redefinition of our $\sigma$ in the asymptotic expansion in  $\varrho$ in order we remove the undesired terms. Again, without loss of generality we will focus on outgoing null geodesics. After performing the change of coordinate \eqref{eqn:hyper_coord} to the Kerr solution in Kerr-Schild coordinates \eqref{eq:Kerr_KS}, we perform the following change of variable
\begin{align}
\sigma &\to \sigma_p+\varrho_p^3 e^{-\sigma_p} \frac{G M (4 \sigma_p-1)}{2 (1+|\zeta_p|^2)^3}\, ,  \qquad \varrho \to \varrho_p-\varrho_p^4 e^{-\sigma_p} \frac{G M (12 \sigma_p+1)}{2 (1+|\zeta_p|^2)^3} \, , \nonumber \\
&\qquad \zeta \to \zeta_p + \varrho_p^4 e^{-2 \sigma_p} \frac{8 i a G M \zeta_p}{(1+|\zeta_p|^2)^4} +\varrho_p^5 e^{-\sigma_p} \frac{G M (12 \sigma_p+1) \zeta_p}{(1+|\zeta_p|^2)^4}\, , \nonumber \\
&\qquad \bar{\zeta} \to \bar{\zeta}_p -\varrho_p^4 e^{-2 \sigma_p} \frac{8 i a G M \bar{\zeta}_p}{(1+|\zeta_p|^2)^4}+ \varrho_p^5 e^{-\sigma_p}\frac{G M (12 \sigma_p+1) \bar{\zeta}_p}{ (1+|\zeta_p|^2)^4} \, ,
\end{align}
to get the metric in Gaussian normal coordinates,
\begin{align}
\hspace{-20pt}&d s_{\text{Kerr},(-)}^2 = e^{2 \sigma_p}  \frac{d \varrho_p^2+d\zeta_p d\bar{\zeta}_p }{\varrho_p^2}-e^{2 \sigma_p} (d\sigma_p)^2 -\varrho_p  e^{\sigma_p} \sigma_p \frac{16 G M \left(2 d \varrho_p^2-d\zeta_p d\bar{\zeta}_p \right)}{(1+|\zeta_p|^2)^3} \nonumber \\
&+ \varrho_p a \frac{48 i d \varrho_p G M (\zeta_p d\bar{\zeta}_p-\bar{\zeta}_p d\zeta_p)}{(1+|\zeta_p|^2)^4} - \varrho_p a^2 e^{-\sigma_p} \frac{16 G M (\zeta_p d\bar{\zeta}_p-\bar{\zeta}_p d\zeta_p)^2}{(1+|\zeta_p|^2)^5}\nonumber \\
&+\varrho^2_p e^{\sigma_p} \sigma_p \frac{96 d \varrho_p G M (\zeta_p d\bar{\zeta}_p+\bar{\zeta}_p d\zeta_p)}{(1+|\zeta_p|^2)^4} + a \varrho^2_p \frac{48 i G M \left(\bar{\zeta}_p^2 (d\zeta_p)^2-\zeta_p^2 (d\bar{\zeta}_p)^2\right)}{(1+|\zeta_p|^2)^5} + \mathcal{O}(\varrho_p^3) .
\end{align}
Upon setting $a = 0$, this gives also the result for the Schwarzschild metric in \eqref{eq:Schw-hyperbslic},
\begin{align}
d s_{\text{Schw},(-)}^2 &= e^{2 \sigma_p} \frac{(d \varrho_p)^2+d\zeta_p d\bar{\zeta}_p }{\varrho_p^2}-e^{2 \sigma_p} (d\sigma_p)^2-\varrho_p \sigma_p e^{\sigma_p} \frac{16 G M \left(2 (d \varrho_p)^2-d\zeta_p d\bar{\zeta}_p \right)}{(1+|\zeta_p|^2)^3} \nonumber \\
&+\varrho_p^2 \sigma_p e^{\sigma_p} \frac{96 G M  (d \varrho_p)(\zeta_p d\bar{\zeta}_p+\bar{\zeta}_p d\zeta_p)}{(1+|\zeta_p|^2)^4} + \mathcal{O}(\varrho_p^3)~ .
\end{align}

It is easy to check that the Ricci scalar of the foliation is still constant and negative on each slice of fixed $\sigma_p$ and we therefore still have an asymptotically AdS$_3$ space (because the redefinitions of the coordinate, despite involving $\sigma$, start at order $\varrho^3$), both for Schwarzschild and Kerr. Using Brown-Henneaux boundary conditions \cite{Brown:1986nw}, our result shows that the asymptotic metric on the foliation is clearly preserved under the action of conformal Killing vectors near the boundary, which suggests the presence a two-dimensional CCFT structure for Schwarzschild and Kerr. Moreover, such Killing vectors can be locally uplifted to the superrotation Killing vectors of asymptotically flat spacetimes \cite{Barnich:2011ct,Ball:2019atb}.

}

\bibliographystyle{JHEP-2}
\bibliography{references}

\end{document}